# First-principles analysis of the Al-rich corner of Al-Li-Cu phase diagram

S. Liu [a], J. S. Wróbel [b], J. LLorca [a,c,1]

[a] IMDEA Materials Institute, C/Eric Kandel 2, 28906 Getafe, Madrid, Spain

[b] Faculty of Materials Science and Engineering, Warsaw University of Technology, ul. Wołoska 141, 02-507 Warsaw, Poland

[c] Department of Materials Science. Polytechnic University of Madrid. E. T. S. de Ingenieros de Caminos. 28040 Madrid, Spain

## Abstract

The phase diagram of Al-Li-Cu system in the Al-rich region was determined by means of first-principles calculations and statistical mechanics. The mixing enthalpies of many configurations for different lattices in the whole Al-Li-Cu system were determined by density functional theory simulations to find the stable phases in the convex hull. They were fitted with a cluster expansion to calculate the free energy of the configurations with different compositions as a function of temperature in the Al-rich region (Al content > 40 at. %) by means of Monte Carlo simulations. It was found that the ground state phases in the Al-rich part of the Al-Li-Cu phase diagram were α-Al, θ' ($Al_2Cu$), δ' ($Al_3Li$), δ (AlLi) and $T_1$ ($Al_6Cu_4Li_3$), while θ'' ($Al_3Cu$), $T_1$' ($Al_2CuLi$) and $Al_3Cu_2Li$ were found on the lowest mixing enthalpy surfaces of their lattices and were metastable. α-Al, δ and $T_1$ are stable phases in the whole temperature range while δ' becomes metastable at very low temperature and θ ($Al_2Cu$) replaces θ' as the stable phase at approximately 550 K due to the vibrational entropic contribution. In addition, the phase diagram in the Al-rich region was built and it was shown in isothermal sections from 100 K to 900 K. They were in good agreement with the limited experimental data in the literature and provided new information regarding the stability, solubility and stoichiometry of the different phases. This information is important to understand the precipitation mechanisms during high temperature aging.

---

[1] Corresponding Author.

Email address: javier.llorca@upm.es , javier.llorca@imdea.org (J. LLorca)

## 1. Introduction

Al-Li-Cu alloys are a class of Al alloys that stand out for structural applications in aerospace. The addition of Li reduces the density and improves the elastic modulus while they present high strength induced by the contribution of Cu and Li in solid solution as well as by the presence of binary and ternary precipitates [1-6]. Obviously, the optimum strength can be achieved through the appropriate thermo-mechanical treatments to attain an optimum distribution of metastable and stable phases [3,5]. Accurate phase diagrams of the Al-rich region are necessary for this purpose.

The most relevant binary precipitates in Al-Li-Cu alloys from the viewpoint of strengthening are θ'' ($Al_3Cu$), θ' ($Al_2Cu$) and δ' ($Al_3Li$) phases [7, 8, 9]. θ'' and θ' are metastable precipitates found in Al-Li-Cu alloys. They appear during ageing at elevated temperature from the supersaturated solid solution (SSS) following the precipitate sequence: SSS → Guinier-Preston (GP) zones → θ'' ($Al_3Cu$) → θ' ($Al_2Cu$) → θ ($Al_2Cu$) [10]. The GP zones are monolayers of Cu atoms parallel to the (001) planes of α-Al matrix formed at the early stages of phase separation [11]. The θ'' ($Al_3Cu$) precipitates are circular disks coherent with α-Al matrix which also grow with the broad face parallel to the (001) planes of the α-Al matrix [12]. The θ' ($Al_2Cu$) precipitates are also disk-shaped precipitates with an orientation relationship $(100)_{\theta'}//(100)_{Al}$, $[001]_{\theta'}//[001]_{Al}$ [12], and they are replaced by the equilibrium θ ($Al_2Cu$) phase when the aging temperature is high enough [12,13]. However, the transformation from θ' ($Al_2Cu$) to θ ($Al_2Cu$) is associated with a large reduction in the yield strength because the large θ precipitates are not strong obstacles to dislocation motion [14].

Another important strengthening phase is δ' ($Al_3Li$), which precipitates homogeneously during the aging of a supersaturated Al-Li-Cu solid solution. The precipitation sequence of δ' ($Al_3Li$) during high temperature aging is: SSS → δ' ($Al_3Li$) → δ (AlLi) [15]. δ' precipitates are spherical and coherent with α-Al matrix [16,17] even when they reach large diameters (300 nm) because of the very small lattice mismatch with α-Al matrix [17,18]. The equilibrium δ (AlLi) phase always precipitates at grain boundaries and its growth is associated with the coarsening and eventual disappearance of δ' ($Al_3Li$) precipitates, which leads to a δ' ($Al_3Li$) precipitate-free zone near grain boundaries [19-22].

Besides the binary phases, the most relevant precipitate in the Al-rich region of the ternary Al-Li-Cu system is the $T_1$ phase. $T_1$ precipitates are reported to have hexagonal structure with stoichiometry $Al_2CuLi$ [23-29] but other stoichiometries have also been reported [30]. They are plate-shaped with an orientation relationship $(0001)_{T1}//(111)_{Al}$, $(1010)_{T1}//(110)_{Al}$, and $(1120)_{T1}//(211)_{Al}$ [31]. Suzuki et al. [32] suggested that precipitation of $T_1$ might be preceded by another metastable precipitate, $T_1'$, and evidence of this mechanism was provided by selected area electron diffraction (SAED) patterns from a 2090 Al-Li-Cu alloy aged at high temperature which included extra reflections associated with a transitional $T_1'$ phase [33]. The structure of the $T_1'$ phase is orthorhombic with a=0.2876nm, b=0.86nm, and c=0.406nm and an orientation relationship $(010)_{T1'}//(011)_{Al}$ and $[001]_{T1'}//[100]_{Al}$.

Depending on the alloy composition and processing conditions, other minor phases, such as $T_2$ and $T_B$, have been reported [34,35]. $T_2$ precipitates have an icosahedral structure with fivefold diffraction symmetry [36] and can reach very large size (up to 2 mm) [36-38]. The stoichiometry of $T_2$ phase is controversial. Some investigations have reported $Al_6CuLi_3$ [39,40] or $Al_5CuLi_3$ [35,40] (also denominated R phase), but other compositions have been found experimentally [41]. The stoichiometry of the $T_B$ phase is generally assumed to be $Al_7Cu_4Li$ [42] and is always considered a metastable phase [43] because it can be dissolved during heat treatments at higher temperature. A reasonable hypothesis to explain the presence of $T_B$ precipitates at low temperature was the influence of strain, which promotes precipitation at high-angle grain boundaries [34, 43].

The currently available phase diagrams of the Al-Li-Cu system are obtained from experimental investigations about phase relations, structure, and thermodynamics [6]. However, the accuracy of the phase diagrams obtained from experiments is not always as good as required because of the difficulties imposed by kinetics and metastability.

Moreover, a systematic experimental study of the Al-Li-Cu phase diagram is hindered by the large number of phases. Thus, it is important to complete the available information from theoretical results provided through the combination of first-principles simulations, statistical mechanics and the cluster expansion (CE) formalism [44-49], a strategy that has proven its potential to determine the thermodynamic properties of multicomponent systems. The CE is formalized as a linear series of cluster basis functions which can be used to determine accurately and efficiently the mixing enthalpy of many configurations in the alloy considered [49-52]. This information can be used to predict the thermodynamic properties of different stable and metastable phases using Monte Carlo (MC) simulations as well as to discover new phases with unknown structure [53].

This strategy is used in this investigation to explore the Al-rich corner of Al-Li-Cu phase diagram. The mixing enthalpies of different configurations for various lattices were determined from first-principles calculations based on density functional theory (DFT) and a CE was fitted for each lattice. They were used to calculate the Gibbs free energy at elevated temperature using MC simulations and to predict the stable and metastable phases as a function of temperature. Moreover, this information was used to predict the unknown structure of $T_1$, $T_B$ and $T_1'$ phases. These results improve our experimental understanding of the Al-Li-Cu phase diagram, an information that is relevant to optimize precipitation heat treatments for this complex ternary alloy.

## 2. Methodology

### 2.1 First-principles calculations

Five different crystal lattices were chosen taking into account to the structures of phases in the Al-rich corner of the Al-Li-Cu system. This selection will be justified in section 3.1. They are face-centered cubic (fcc), body-centered cubic (bcc), hexagonal, body-centered tetragonal (bct) and cubic (corresponding to that of $CaF_2$) (Fig. S1 in the Supplementary Information). Symmetrically distinct configurations with Al, Cu, Li atoms randomly placed on these lattice sites were generated. The atomic positions, lattice parameters and angles of each $Al_{1-x-y}Cu_xLi_y$ configuration were fully relaxed at pressure P=0 using DFT calculations with Quantum Espresso [54,55]. The electron exchange-correlation was described using the generalized gradient approximation with the Perdew-Burke-Ernzerhof exchange-correlation functional and ultrasoft pseudopotentials [56,57] and an energy cut-off of 114 Ry. The Brillouin zone was sampled using a Monkhorst-Pack grid with a density of 40 points/Å$^{-1}$.

The mixing enthalpy per atom of each $Al_{1-x-y}Cu_xLi_y$ configuration with respect to those of pure Al, Cu, Li in the corresponding lattice system *s*, $H_s^{mix}(Al_{1-x}Cu_xLi_y)$, was calculated according to

$$H_s^{mix}(Al_{1-x}Cu_xLi_y) = E_s(Al_{1-x}Cu_xLi_y) - (1-x-y)E_s(Al) - xE_s(Cu) - yE_s(Li) \quad (1)$$

where $E_s(Al_{1-x}Cu_xLi_y)$ is the energy per atom of the configuration after relaxation and $E_s(Al)$, $E_s(Cu)$, $E_s(Li)$ stand for the relaxed energies per atom of pure Al, Cu, Li in the corresponding lattice system, respectively. It should be noted that the mixing enthalpy per atom of eq. (1) will be used to fit the CE but it does not correspond to the actual formation enthalpy, $H^f$, that is expressed as

$$H^f(Al_{1-x}Cu_xLi_y) = E_s(Al_{1-x}Cu_xLi_y) - (1-x-y)E(Al) - xE(Cu) - yE(Li) \quad (2)$$

where $E(Al)$, $E(Cu)$ and $E(Li)$ stand for the relaxed energy per atom of Al, Cu and Li in their equilibrium structure, namely fcc for Al and Cu, and bcc for Li.

The lattice distortion of each configuration after relaxation was determined to assess whether the lattice symmetry changed. Following the criteria reported in the literature [58,59], only configurations whose distortion was below 10% were considered to keep the original lattice symmetry and used to fit the CE for each lattice [60]. 1288 configurations with up to 6 atoms per unit cell were calculated with the fcc lattice and only 799 were used to fit CE form this lattice. 1208 configurations with up to 13 atoms per unit cell were calculated with the bcc lattice and 666 were used to fit the CE. 1021 configurations with up to 26 atoms per unit cell were calculated in the hexagonal system and 795 were used to fit the CE. 468 with up to 12 atoms per unit cell were calculated in the bct system and 229 were used to fit the CE. Finally, 521 configurations with up to 48 atoms per unit cell were calculated in the cubic $CaF_2$ system and 289 were initially used to fit the CE. Later, another 100 configurations with up to 96 atoms per unit cell were further added to fit the CE.

**2.2 Cluster expansions for different structures**

The enthalpies of mixing calculated by DFT for each lattice structure were fitted by the CE algorithm following the formalism in the ATAT package [61]:

$$H_s^{mix}(\vec{\sigma}) = \sum_{\omega} m_{\omega} V_{\omega} \langle \varphi(\vec{\sigma}) \rangle_{\omega} \tag{3}$$

where the summation indicates all distinct clusters $\omega$ under symmetry operation. $m_\omega$ and $V_\omega$ stand, respectively, for the multiplicity factor and effective cluster interaction (ECI) coefficient of each cluster. $\langle \varphi(\vec{\sigma}) \rangle_{\omega}$ is the cluster function of $\omega$ defined as a product of orthogonal point functions $\gamma_j(\sigma_i)$ over all sites included in $\omega$ as:

$$\varphi^{(d)}(\vec{\sigma}) = \gamma_{j_1}(\sigma_1)\gamma_{j_2}(\sigma_2) \cdots \gamma_{j_\omega}(\sigma_\omega) \tag{4}$$

where $\sigma_i = (0, 1, 2, \ldots, M-1)$ indicates which type of atom sits on lattice site $i$ in a M-component system. $(d)$ is the decoration [62] of the cluster $\omega$ by point functions $\gamma_{j_1}$ to $\gamma_{j_\omega}$, where $j_i = (0, 1, 2, \ldots, M-1)$ has the similar meaning as $\sigma_i$, indicating which type of atom sits on lattice site *i* belong to cluster $\omega$. The cluster functions for each two clusters α and β should satisfy $\langle \varphi(\vec{\sigma})_\alpha, \varphi(\vec{\sigma})_\beta \rangle = 0$ if they differ and $\langle \varphi(\vec{\sigma})_\alpha, \varphi(\vec{\sigma})_\beta \rangle = 1$ if they are identical. This is achieved with point functions $\gamma_j(\sigma_i)$ expressed as [61]:

$$\gamma_{j_i}(\sigma_i) = \begin{cases} 1 & \text{if } j = 0 \\ -\cos\left(2\pi \left[\frac{j_i}{2}\right] \frac{\sigma_i}{K}\right) & \text{if } j > 0 \text{ and odd} \\ -\sin\left(2\pi \left[\frac{j_i}{2}\right] \frac{\sigma_i}{K}\right) & \text{if } j > 0 \text{ and even} \end{cases} \tag{5}$$

where $\left[\frac{j_i}{2}\right]$ denotes the 'round down' operation. The point functions with j>0 and odd are denominated as function '0' while those with j>0 and even are named as function '1' in the description of the CEs for each lattice in the Supplementary Information.

### 2.3 Monte Carlo simulations

The Helmholtz free energy $F$ at a given temperature $T$ can be calculated as

$$F = H^f(T) - TS_{conf}(T) \tag{6}$$

where $H^f(T)$ is the formation enthalpy and $S_{conf}(T)$ is the configurational entropy at $T$. $H^f(T)$ is obtained from the mixing enthalpy $H_s^{mix}$(T) that is computed using eq. (3) by averaging the values from all MC simulations steps in the accumulation stage for a given temperature $T$. $S_{conf}(T)$ can be determined from the integration of the configurational contribution to the specific heat $C_{conf}(T)$ from 0 K to $T$ according to

$$S_{conf}(T) = \int_0^T \frac{C_{conf}(T')}{T'} dT' \tag{7}$$

where $C_{conf}(T)$ can be calculated from the fluctuations of mixing enthalpy at a given temperature according to [62,63]:

$$C_{conf}(T) = \frac{\langle H_s^{mix}(T)^2\rangle - \langle H_s^{mix}(T)\rangle^2}{T^2} \tag{8}$$

where $\langle H_s^{mix}(T)^2\rangle$ and $\langle H_s^{mix}(T)\rangle^2$ stand for the mean enthalpy and the mean-square mixing enthalpies, respectively, which are computed by averaging over all the MC passes at the accumulation stage for a given temperature. In this work, MC simulations in canonical ensemble were performed using a simulation box containing 20×20×20 unit cells for each composition in Al-Li-Cu alloy. 1000 MC passes were allowed at each temperature to reach equilibrium and 2000 passes were performed to compute the mean and the mean-square mixing enthalpies. In order to get the integration of the configurational entropy through eq. (7), MC simulations from 1000K to 10K with a temperature step of 10K at each composition were performed. In the case of solid-state transformation, the difference between Helmholtz free energy and Gibbs free energy can be neglected and, thus, the Helmholtz free energy from eq. (6) can be used to construct the phase diagram [64].

## 3. Results and discussion

### 3.1 Phases of interest

The stable and metastable phases in the Al-rich corner of Al-Li-Cu phase diagram - according to the information in the literature [2] - are depicted in Fig. 1. Al has a fcc lattice structure (Fig. 2(a)) and the two nearest phases are θ'' ($Al_3Cu$) and δ' ($Al_3Li$). θ'' ($Al_3Cu$) also presents a fcc lattice, with three Al(001) layers sandwiched between two Cu(001) monolayers (Figs. 2(b)). δ' ($Al_3Li$) has a $L1_2$ (ordered fcc) superlattice crystal structure and different elements occupy the corner and face center sites, respectively (Figs. 2(c)). Therefore, Al, θ'' ($Al_3Cu$) and δ' ($Al_3Li$) are plotted with the same red color in Fig. 1.

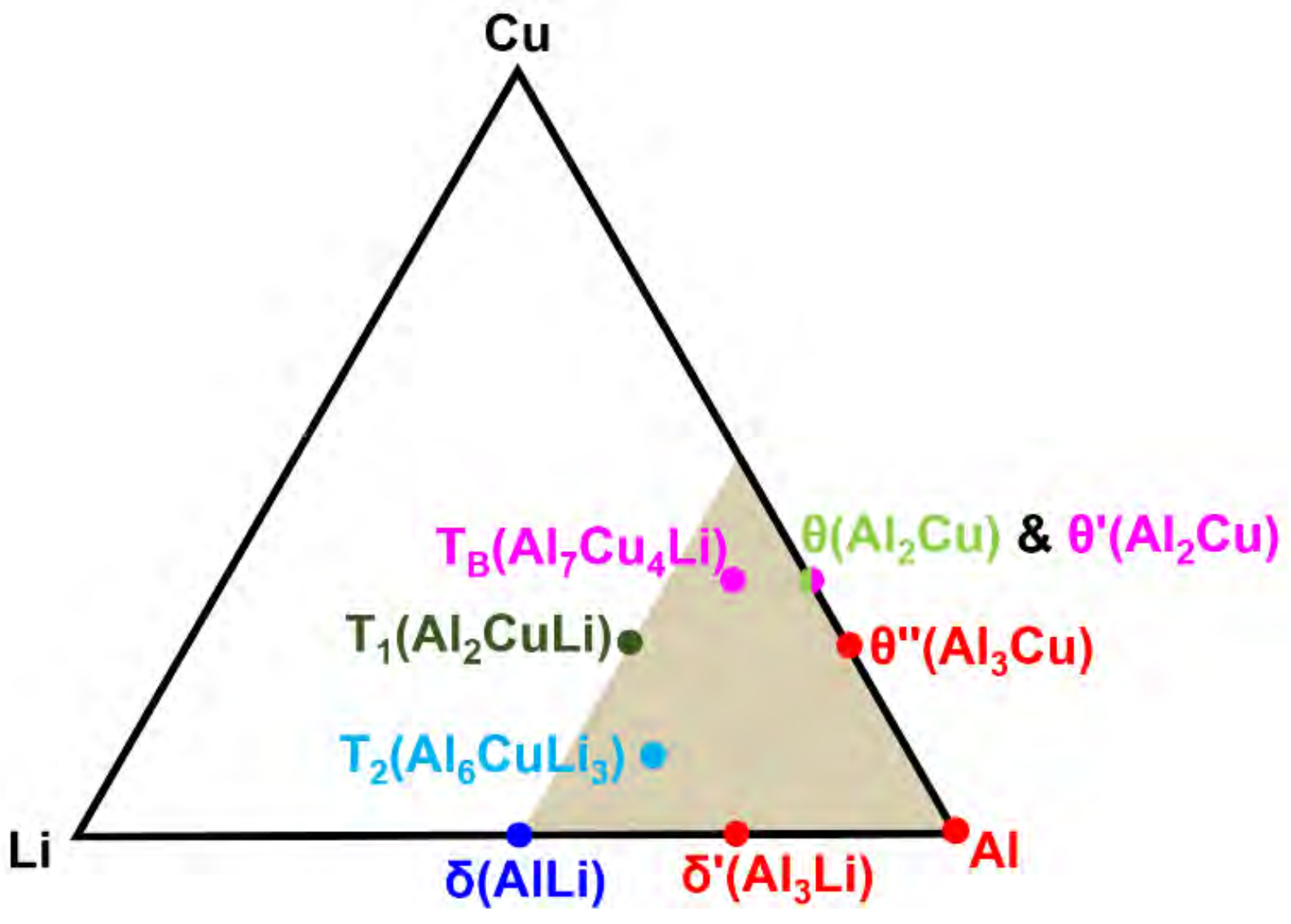

**Fig. 1.** Stable and metastable phases in the Al-rich corner (shadowed area) of Al-Li-Cu ternary system. They are indicated with different colors according to their lattice. Red stands for fcc (Al, θ'' and δ'), navy blue for bcc (δ), green for bct (θ), pink for cubic $CaF_2$ (θ' and $T_B$), olive green for hexagonal ($T_1$) and skyblue for icosahedral ($T_2$). θ and θ' are superposed in this representation because they have the same stoichiometry.

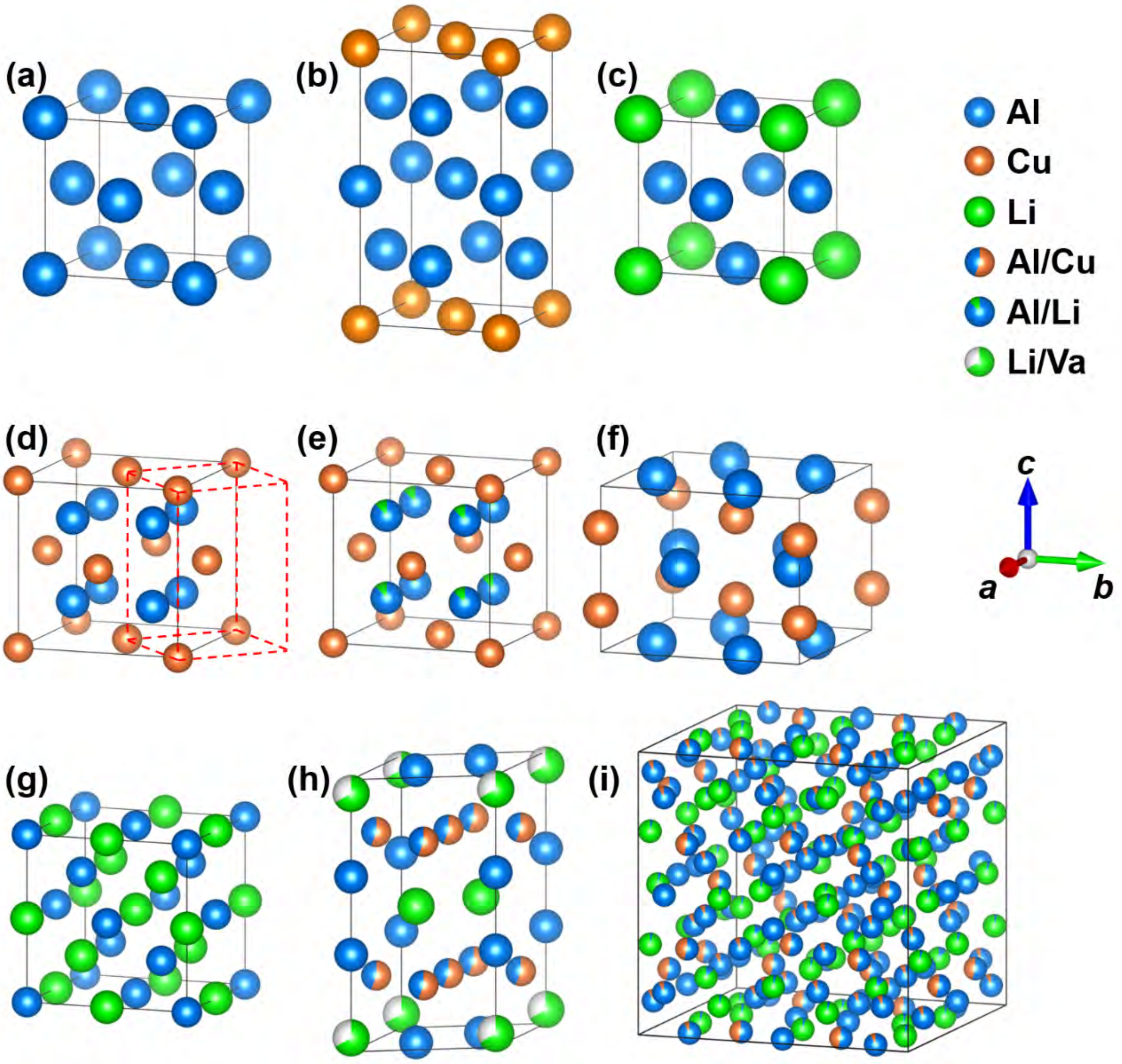

**Fig. 2.** Structure of the different phases in the Al-rich corner of the Al-Li-Cu ternary system. (a) Al. (b) θ'' ($Al_3Cu$). (c) δ' ($Al_3Li$). (d) θ' ($Al_2Cu$). (e) $T_B$ ($Al_7Cu_4Li$). (f) θ ($Al_2Cu$). (g) δ (AlLi). (h) $T_1$ ($Al_2CuLi$). (i) $T_2$ ($Al_6CuLi_3$). Al atoms are blue, Cu atoms orange, Li atoms green. Positions that can be filled with either Al or Cu atoms are plotted half blue and half orange, those that can be filled with either Al or Li atom are plotted half blue and half green while those that can be occupied by Li or vacancy are plotted half green and half grey.

θ' ($Al_2Cu$) is always considered to have a bct structure [11], that is indicated by the dashed lines in Fig. 2(d). Nevertheless, this structure can also be recognized as a cubic $CaF_2$ structure made up by a unit cell whose dimensions are $\sqrt{2} \times \sqrt{2} \times 1$ times those of the bct cell (Fig. 2(d)). In fact, Cu is coordinated to eight Al atoms and each Al atom is surrounded by four Cu atoms, as in the $CaF_2$ structure. $T_B$ ($Al_7Cu_4Li$) (Fig. 2(e)) also shares the same lattice as θ' ($Al_2Cu$) and the only difference is that the Al sites in θ' ($Al_2Cu$) are partially occupied by Li in $T_B$ ($Al_7Cu_4Li$). For the sake of clarity, the sites that can be occupied by Al or Li in $T_B$ ($Al_7Cu_4Li$) are represented using bicolor atoms.

θ ($Al_2Cu$) is another bct structure, which is depicted in Fig. 2(f). δ (AlLi) presents a bcc lattice made up of two interpenetrating diamond sublattices of Al and Li, respectively (Fig. 2(g)). Finally, $T_1$ ($Al_2CuLi$) and $T_2$ ($Al_6CuLi_3$) present hexagonal and icosahedral structures, respectively, that are depicted in Figs. 2(h) and (i). Similar as $T_B$ ($Al_7Cu_4Li$), some lattice sites may be occupied by two different elements and they are indicated by bicolor atoms. The grey color in some atoms of $T_1$ ($Al_2CuLi$) indicates that these sites may be occupied by a vacancy.

### 3.2 Phase stability in each system

The mixing enthalpies of configurations generated from the fcc lattice that do not degenerate to other lattice after relaxation are depicted in Fig. 3(a). They can be used to build the convex hull of this lattice system, which is defined by the most stable configurations in the fcc Al-Li-Cu system indicated in Fig. 3(b). θ'' ($Al_3Cu$) and δ' ($Al_3Li$) are the only configurations on the convex hull near the Al-rich corner and the Al content is $\leq 0.5$ in the rest of the configurations on the fcc convex hull. The information in Fig. 3(a) was used to fit the ECI coefficients of a CE formalism that contains 1 empty cluster, 2 point clusters, 24 pair clusters and 84 triplet clusters. The cross-validation (CV) score of the fcc CE was equal to 0.011eV/atom. The detailed information of the CE, including decoration, multiplicity as well as the actual values of the ECI coefficients of each cluster can be found in Table S1 in the Supplementary Information.

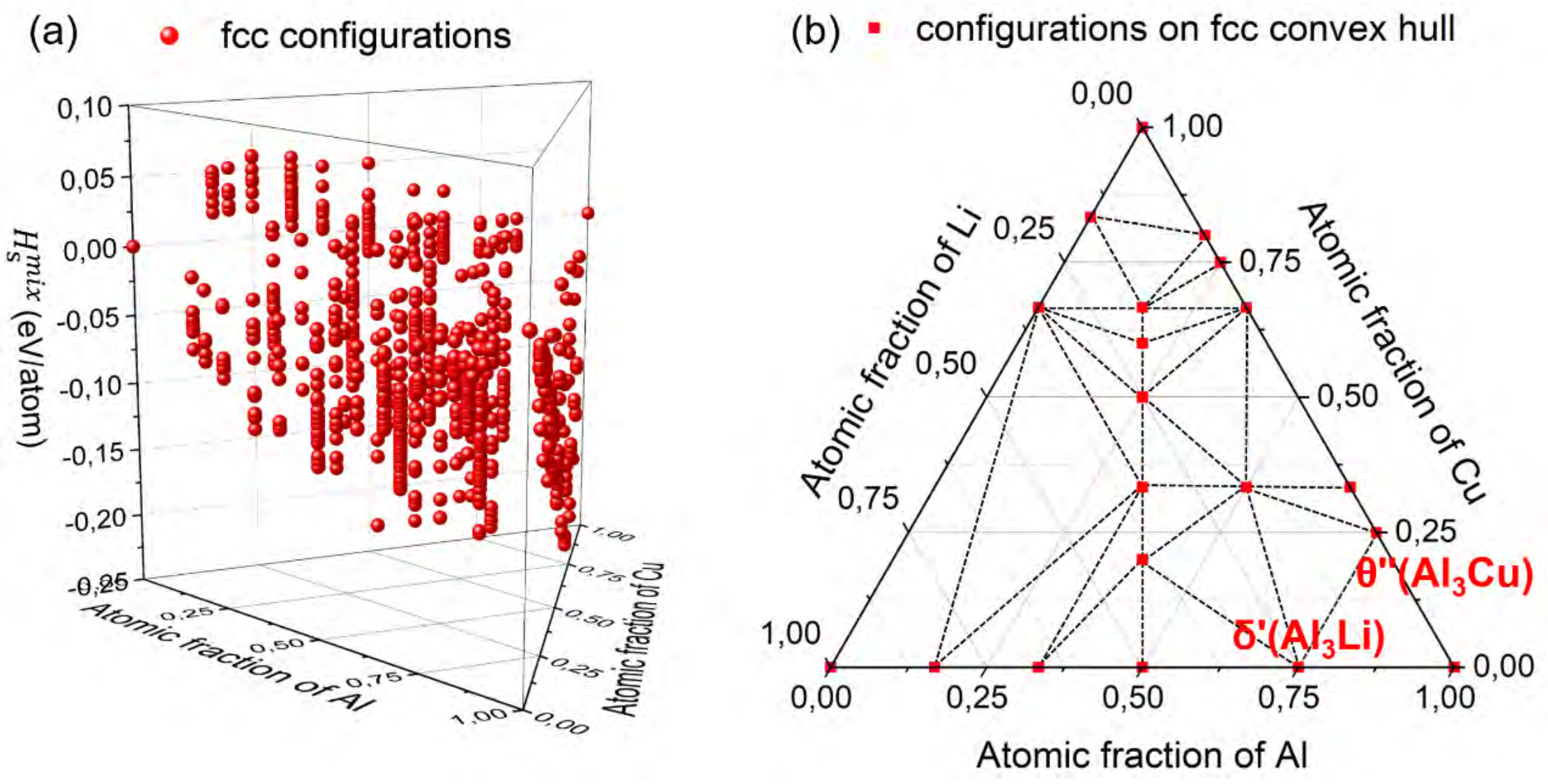

**Fig. 3.** (a) Mixing enthalpies of configurations with fcc lattice with respect to fcc Al, Cu, Li according to eq. (1). (b) Configurations on the fcc convex hull.

The mixing enthalpies of the configurations generated from the bcc lattice that remain with the structure after relaxation are depicted in Fig. 4(a). The configurations with minimum mixing enthalpy that define the convex hull of bcc system are indicated in Fig. 4(b). As expected, δ (AlLi) is on the convex hull because it is known to be a stable phase in Al-Li-Cu and Al-Li alloys. AlCu is also on the bcc convex hull but has never been reported experimentally, to the authors' knowledge. The other configuration on the convex hull nearest to the Al-rich corner is $Al_2CuLi$, which has the same structure as the transitional $T_1$' phase observed in [33] (Fig. 5). The lattice parameters of $Al_2CuLi$, obtained from the DFT calculations, are a = 0.299 nm, b = 0.846 nm and c = 0.423 nm, which are also close to those measured in [33]. Other phases (such as $Al_4Cu_9$, $AlLi_2$, $Al_2Li_3$) are also found on the convex hull of bcc system but they are away from the Al-rich corner and are not further investigated in this paper. The ECI coefficients of a CE formalism were fitted from the information in Fig. 4(a). They include 1 empty cluster, 2 point clusters, 24 pair clusters, 84 triplet clusters and 100 quadruplet clusters. The CV score of the bcc CE was equal to 0.017eV/atom. The detailed information of the CE as well as the actual values of the ECI coefficients of each cluster can be found in Table S2 in the Supplementary Information.

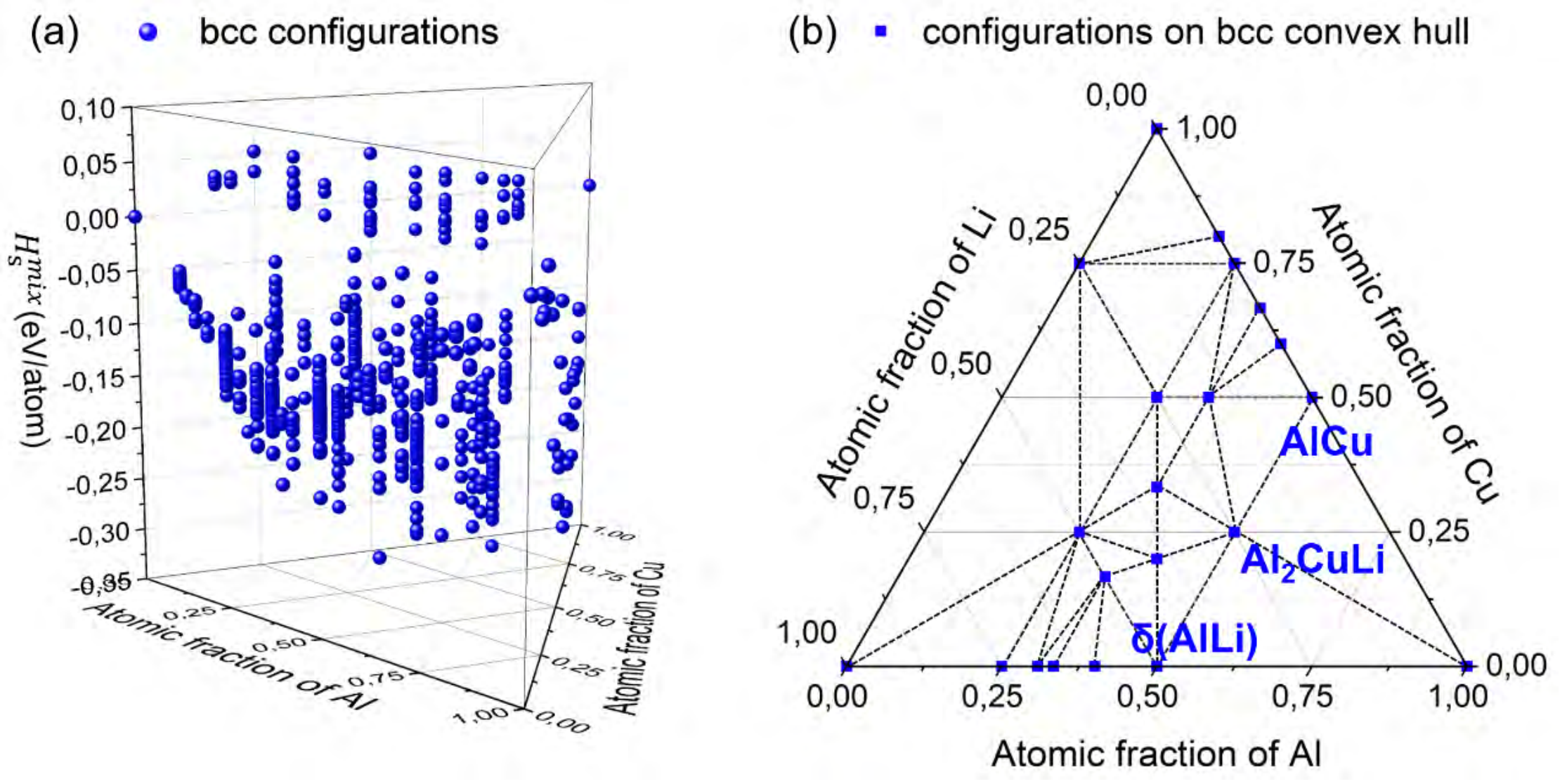

**Fig. 4.** (a) Mixing enthalpies of configurations with bcc lattice with respect to bcc Al, Cu, Li according to eq. (1). (b) Configurations on the bcc convex hull.

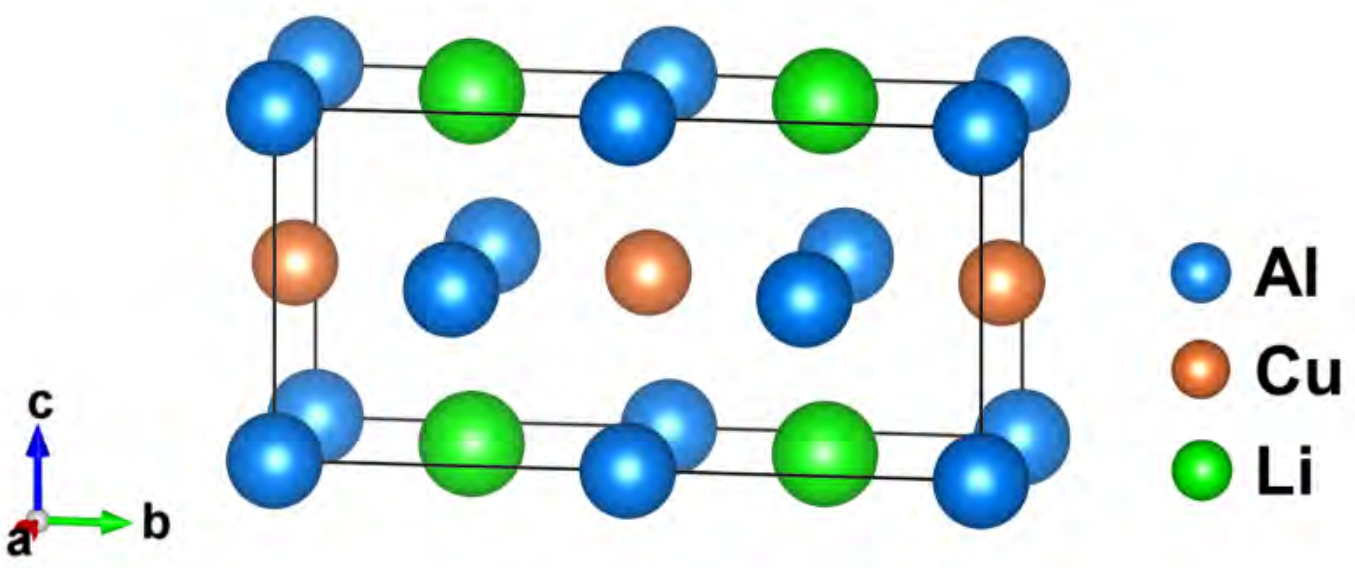

**Fig. 5.** Structure of $T_1$' ($Al_2CuLi$) phase.

The mixing enthalpies of the configurations generated from the $CaF_2$ lattice that remained with this structure after relaxation are depicted in Fig. 6(a). The configurations on the convex hull of the $CaF_2$ system are indicated in Fig. 6(b). θ' ($Al_2Cu$) is one of the configurations on the convex hull nearest to Al and another nearby configuration is $Al_3Cu_2Li$, whose structure is shown in Fig. 7(a). $T_B$ ($Al_7Cu_4Li$), which should be found between θ' ($Al_2Cu$) and $Al_3Cu_2Li$, does not appear on the convex hull. The CE formalism obtained for this structure was used to calculate the mixing enthalpy of other configurations in this area but none of them was on the convex hull. Thus, the mixing enthalpy of another 100 configurations (including 30 configurations with the stoichiometry of $Al_7Cu_4Li$) was added to the previous configurations to fit the coefficients of a new CE but no new configurations with the stoichiometry of $T_B$ were found in the convex hull. The $Al_7Cu_4Li$ configuration with the lowest mixing enthalpy is depicted in Fig. 7(b) and corresponds to the smallest cell with this lattice. The lattice parameters obtained by DFT calculations are a = b = c = 0.573 nm. The distance to the convex hull of the $CaF_2$ system is 0.06 eV/atom. The CE of $CaF_2$ lattice contains 1 empty cluster, 4 point clusters, 47 pair clusters, and 34 triplet clusters with a CV score of 0.021eV/atom. The detailed information of the CE as well as the actual values of the ECI coefficients of each cluster can be found in Table S3 in the Supplementary Information.

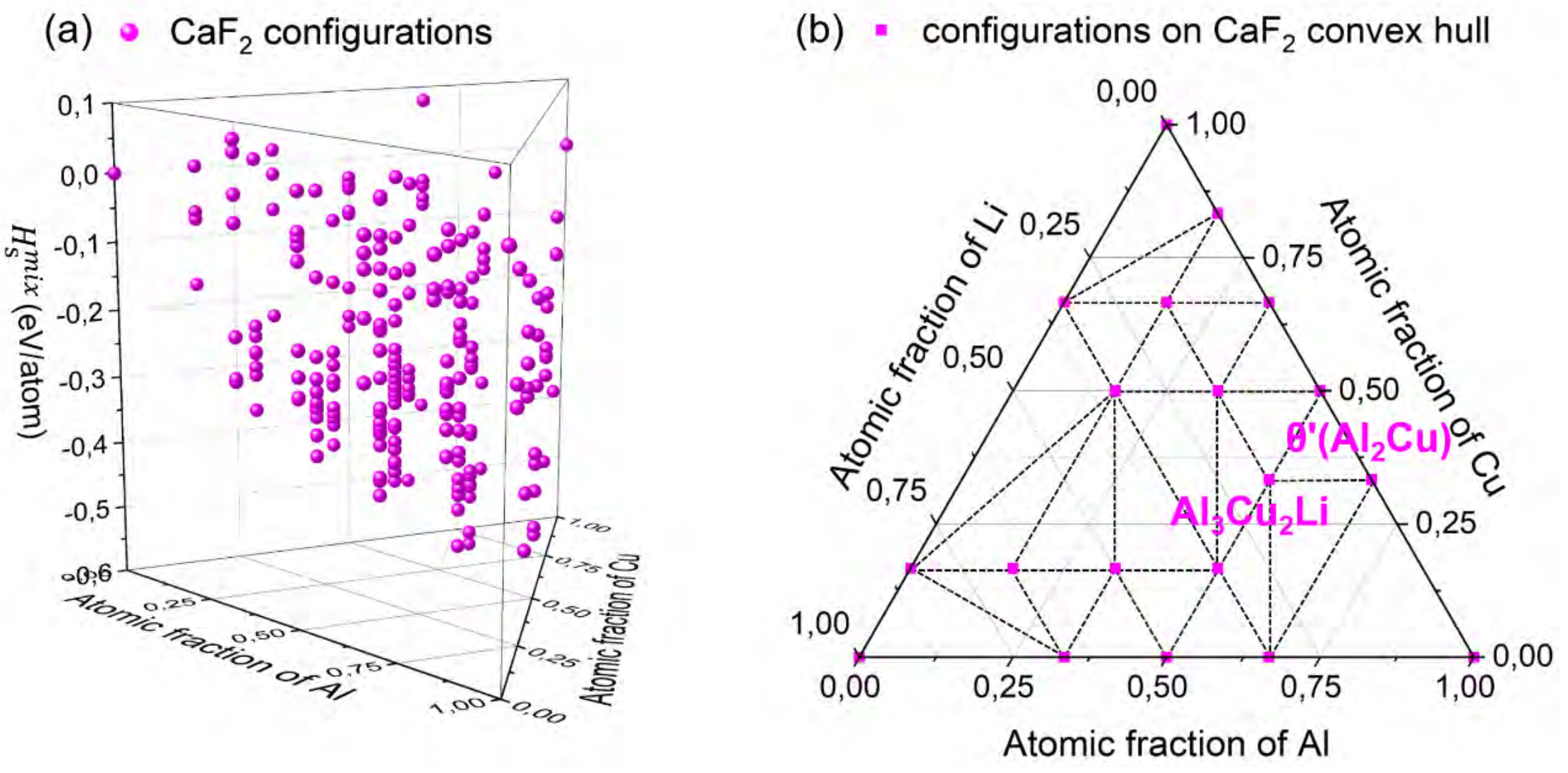

**Fig. 6.** (a) Mixing enthalpies of configurations with $CaF_2$ lattice with respect to Al, Cu, Li with $CaF_2$ lattice according to eq. (1). (b) Configurations on the $CaF_2$ convex hull.

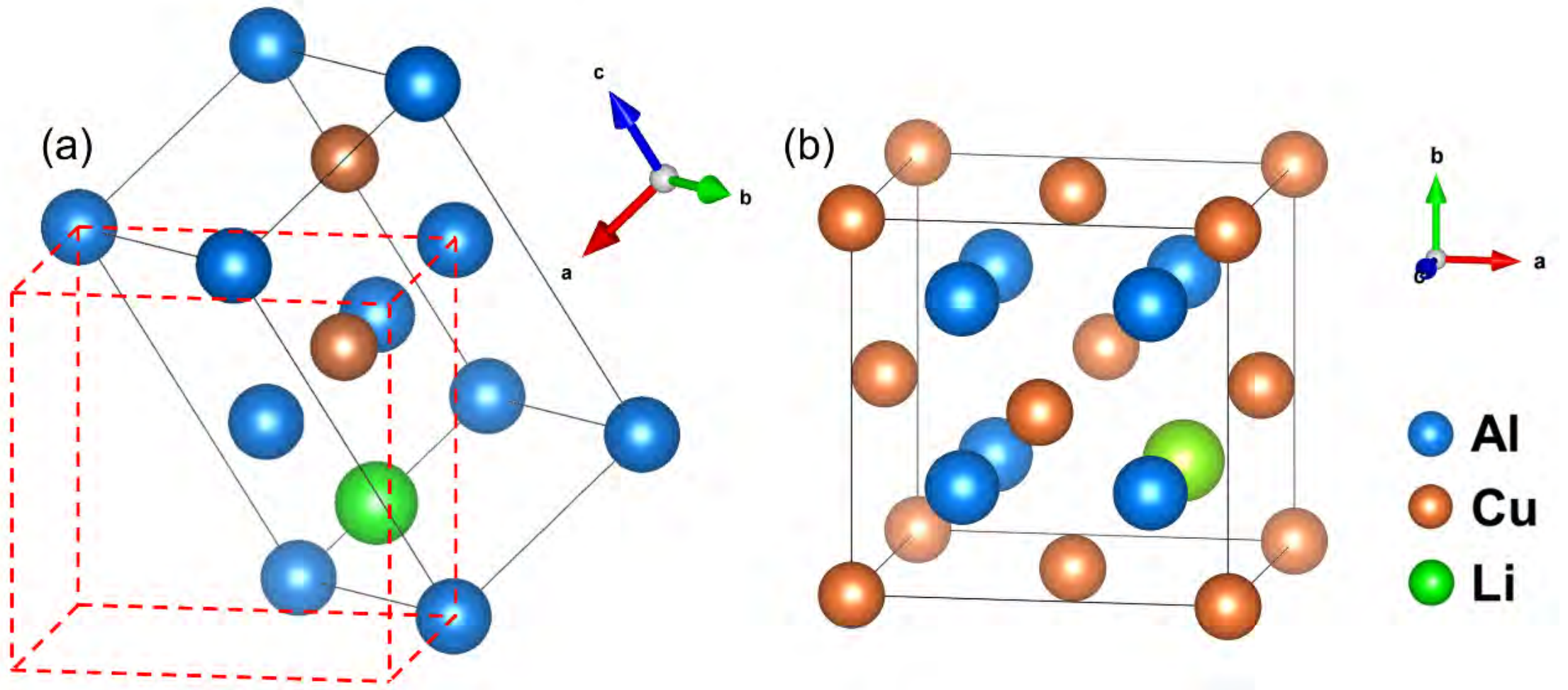

**Fig. 7.** (a) Structure of $Al_3Cu_2Li$, with the $CaF_2$ lattice indicated by the dashed lines; (b) Structure of $T_B$ ($Al_7Cu_4Li$) phase with the lowest energy.

The mixing enthalpies of the configurations generated from the bct lattice that did not degenerate to other symmetry after relaxation are depicted in Fig. 8(a). The configurations on the convex hull of bct system are indicated in Fig. 8(b). θ ($Al_2Cu$) is the only phase on the convex hull in the Al-rich corner. The CE of bct lattice obtained from Fig. 8(a) contains 1 empty cluster, 4 point clusters, 59 pair clusters and 40 triplet clusters, with a CV score of 0.061eV/atom. The detailed information of the CE as well as the actual values of the ECI coefficients for each cluster can be found in Table S4 in the Supplementary Information.

Finally, the hexagonal structure was explored. According to the structure of $T_1$ ($Al_2CuLi$) in Fig. 1(h), some lattice sites can be occupied by either Al or Cu, or by Li or a vacancy. Thus, hexagonal configurations within the possible compositions indicated in Fig. 1(h) were generated and the mixing enthalpies were calculated after relaxation. The mixing enthalpies of the ones that remain hexagonal after relaxation are depicted in Fig. 9(a). The configurations on the hexagonal convex hull are indicated in Fig. 9(b). Instead of $Al_2CuLi$, the configuration on the convex hull that is nearest to the Al-rich corner has the $Al_6Cu_4Li_3$ stoichiometry. According to the mixing enthalpies displayed in Fig. 9(a), a CE with a CV score of 0.054eV/atom was fitted for the hexagonal lattice, which contains 1 empty cluster, 1 point clusters, 27 pair clusters and 19 triplet clusters. The detailed information of the CE as well as the actual values of the ECI of each cluster can be found in Table S5 in the Supplementary Information.

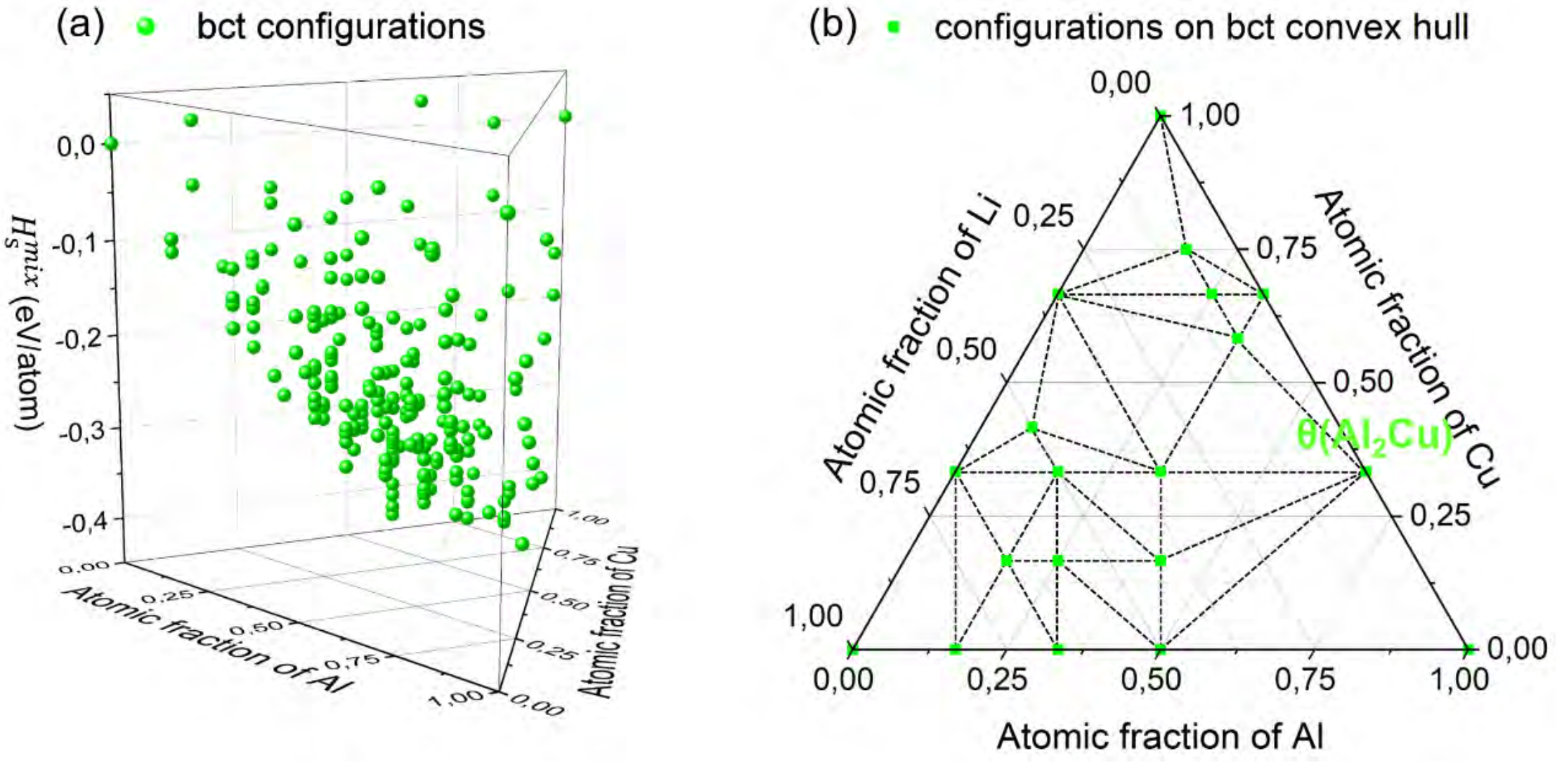

**Fig. 8.** (a) Mixing enthalpies of configurations with bct lattice relative to bcc Al, Cu, Li according to eq. (1). (b) Configurations on the bct convex hull.

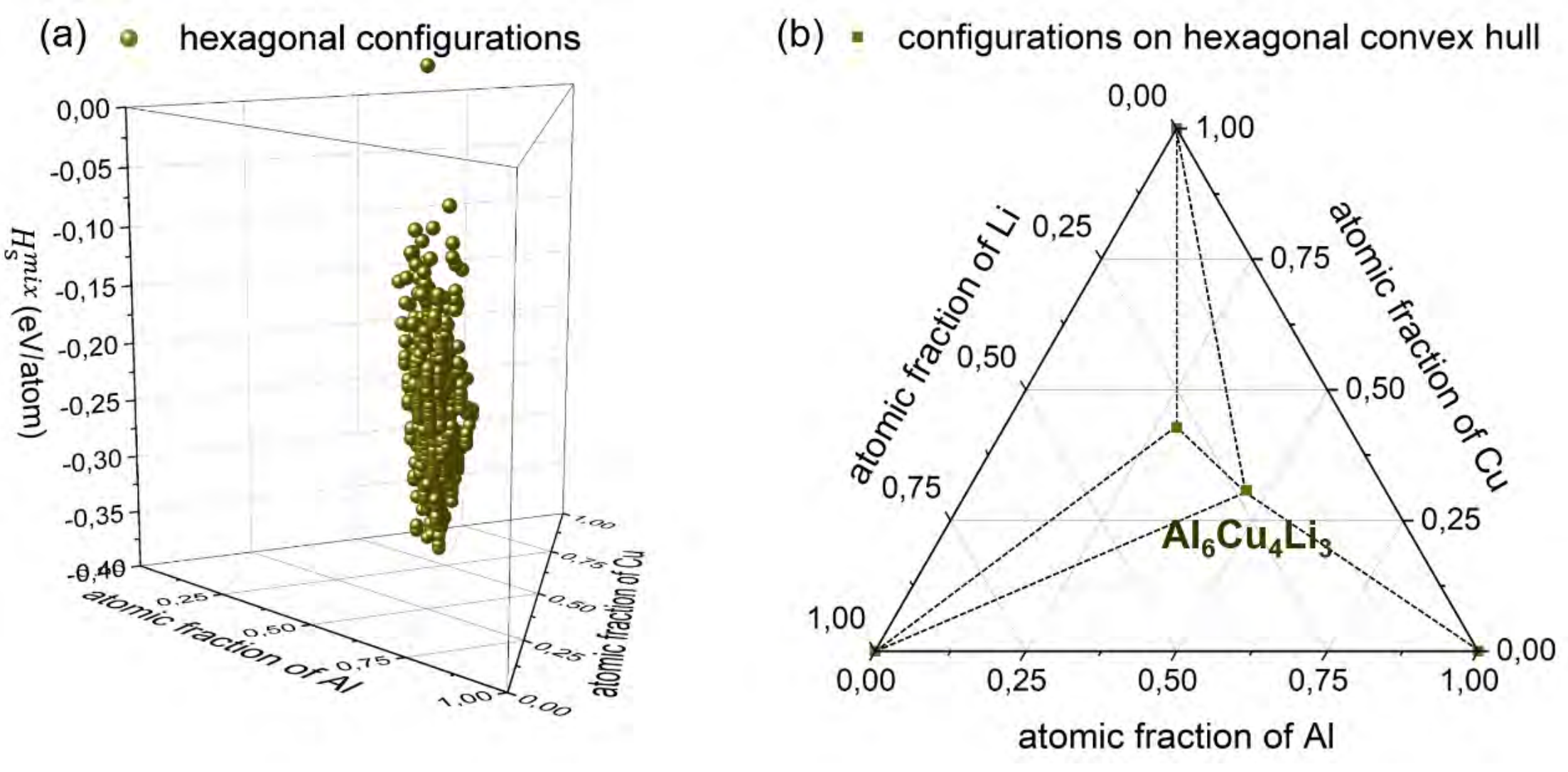

**Fig. 9.** (a) Mixing enthalpies of configurations with hexagonal lattice relative to hexagonal Al, Cu, Li according to eq. (1). (b) Configurations on the hexagonal convex hull.

### 3.3 Phase stability of the Al-Li-Cu system at 0K

The formation enthalpies of the configurations with the five different lattices were compared using the relaxed energies of fcc Al, Cu, and bcc Li as reference, according to eq. (2), and they are plotted in Fig. 10(a). The convex hull surfaces of the different lattices are depicted in Fig. 10(b). They overlap and only the configurations on the lowest surfaces are possible ground state phases. Therefore, the overlapped convex hulls are viewed from the bottom in Fig. 10(c), which only shows the lowest surface and indicates the lattices that provide the minimum energy configurations as a function of the alloy composition. The possible ground state phases are marked in Fig. 10(d) by dots in different colors. In the Al-rich corner (Figs. 10(c) and (d)), they are three phases in the red fcc region, namely pure Al, θ'' ($Al_3Cu$) and δ' ($Al_3Li$), two phases in the navy blue bcc region, namely δ (AlLi) and $T_1$' ($Al_2CuLi$), and two phases in the pink $CaF_2$ region, namely θ' ($Al_2Cu$) and $Al_3Cu_2Li$. No one phase with a bct lattice is found on the lowest surface because the only possible one in this region is θ ($Al_2Cu$) and it has a slightly higher formation enthalpy (0.012 eV/atom) than θ' ($Al_2Cu$) at 0 K. Finally, the new predicted $T_1$ ($Al_6Cu_4Li_3$) phase can be found in the olive region corresponding to the hexagonal lattice.

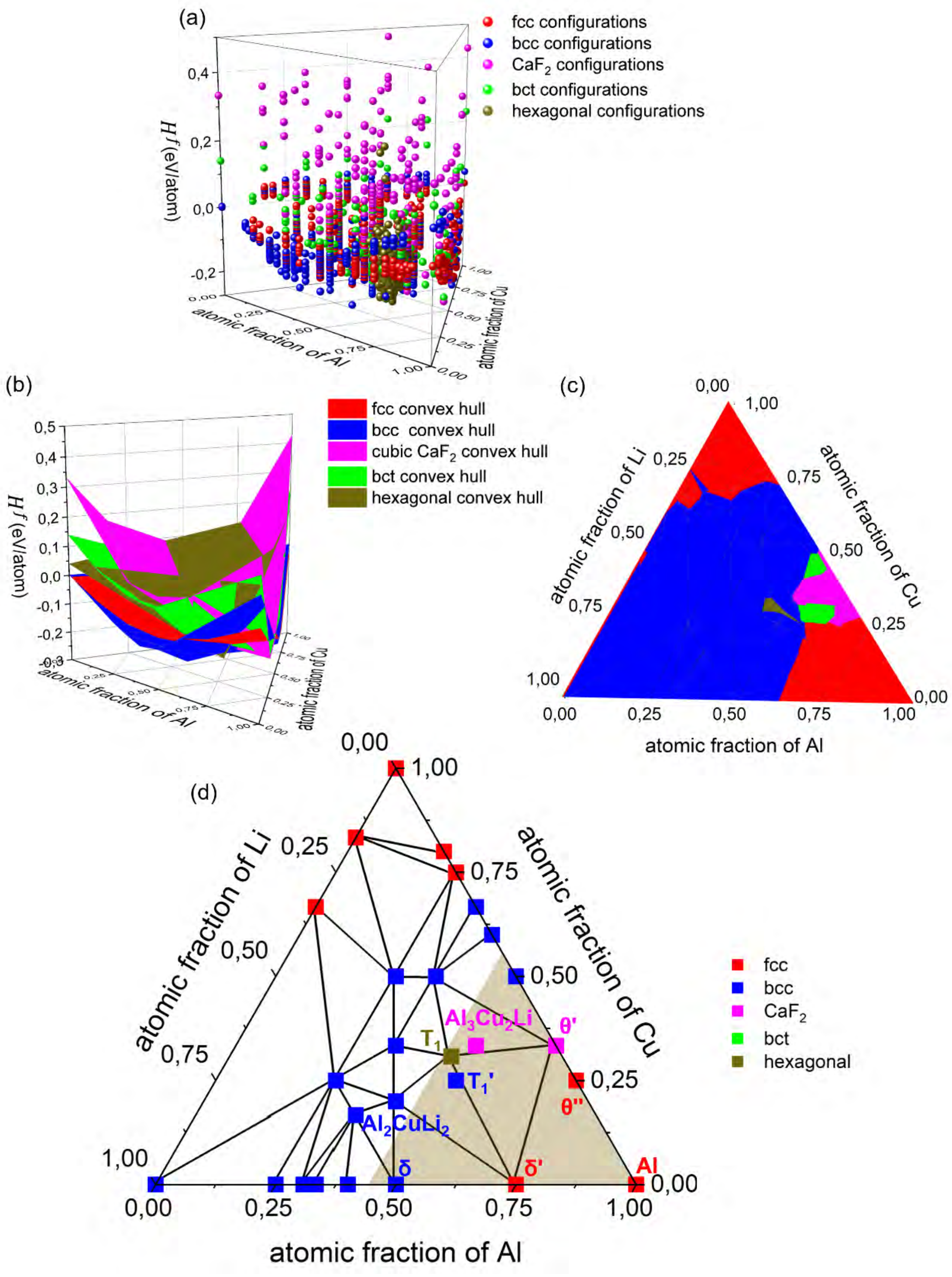

**Fig. 10.** (a) Formation enthalpies of configurations with different lattices relative to fcc Al and Cu, and bcc Li. The $T_2$ ($Al_6CuLi_3$) phase (which is normally formed from melt [36,41] and whose formation enthalpy cannot be assessed with the CE strategy because if the large unit cell, Fig. 2(i)), is not included. (b) Convex hull surfaces of the different lattices. (c) Bottom view of (b). (d) Global convex hull of the Al-Li-Cu system.

According to the convex hull theory, the global convex hull of a ternary system is made up by phases whose formation enthalpy is lower than that of any other phase for a given composition (regardless of the structure) or by a linear combination of phases at that composition. Thus, the ground state phases are among those mentioned above. The global convex hull of Al-Li-Cu - that corresponds to the phase diagram at 0K - is represented by the connecting lines in Fig. 10(d) whereas the aforementioned ones that are not connected by lines are metastable. The stable phases in the Al-rich corner at 0 K are Al, θ' ($Al_2Cu$), δ' ($Al_3Li$), δ (AlLi) and $T_1$ ($Al_6Cu_4Li_3$). The other phases that appear on the lowest mixing enthalpy surfaces of the different lattices, namely θ'' ($Al_3Cu$), $T_1$' ($Al_2CuLi$) and $Al_3Cu_2Li$, are metastable. Their distances to the global convex hull are 0.036 eV/atom for θ'' ($Al_3Cu$), 0.009 eV/atom for $T_1$' ($Al_2CuLi$) and 0.026 eV/atom for $Al_3Cu_2Li$. We should notice that, even the configurations with bct lattice are the ones with minimum energy in the two green regions in Fig. 10(c), they are above the global convex hull that does not include any ground state phases with bct structure.

### 3.4 Phase diagram at elevated temperatures

The CE of each lattice structure can be used to predict the thermodynamic properties of the corresponding structures with this lattice following the methodology presented in section 2.3. Therefore, the compositional space of the Al-Li-Cu ternary system for each lattice was systematically explored in the Al-rich zone (Al > 40 at. %) using an interval of 5 at. % for each one of the constituents. Moreover, additional compositions were explored at intervals of 1 at. % near the convex hull phases. MC simulations at each composition were performed using a simulation box containing 20×20×20 unit cells. The free energy of each composition was calculated as indicated in section 2.3, and the free energy surfaces at each temperature for each lattice system were obtained. Similar to the results in Fig. 10(b), only the lowest free energy surfaces provide useful information to determine the stable phases at each temperature. The boundaries between different stable phase were obtained by the common tangent construction.

The results of these calculations are depicted in isothermal sections of the Al-Li-Cu phase diagram in the Al-rich region, which plotted in Figs. 11 and Fig. S2 in the Supplementary Information from 100K to 900K at intervals of 100K. They show triangular regions plotted in different colors which stand for the stability region for the three different phases at the corners of each triangle. They are α-Al, θ' ($Al_2Cu$) and $T_1$ (blue triangle), α-Al, $T_1$ and $Al_2CuLi_2$ (yellow triangle), and α-Al, $Al_2CuLi_2$ and δ (AlLi) (green triangle), as indicated in Fig. 11(a). In addition, the lines or white triangles between each two three-phase equilibrium regions are the two-phase regions. Finally, the lines or white regions connecting δ (AlLi) and α-Al as well as θ' ($Al_2Cu$) and α-Al are also two-phase regions for the corresponding binary phases. The white polygonal area on the Al corner (clearly visible at temperatures ≥ 500 K) indicates the region in which Cu and Li are in solid solution within the α-Al matrix.

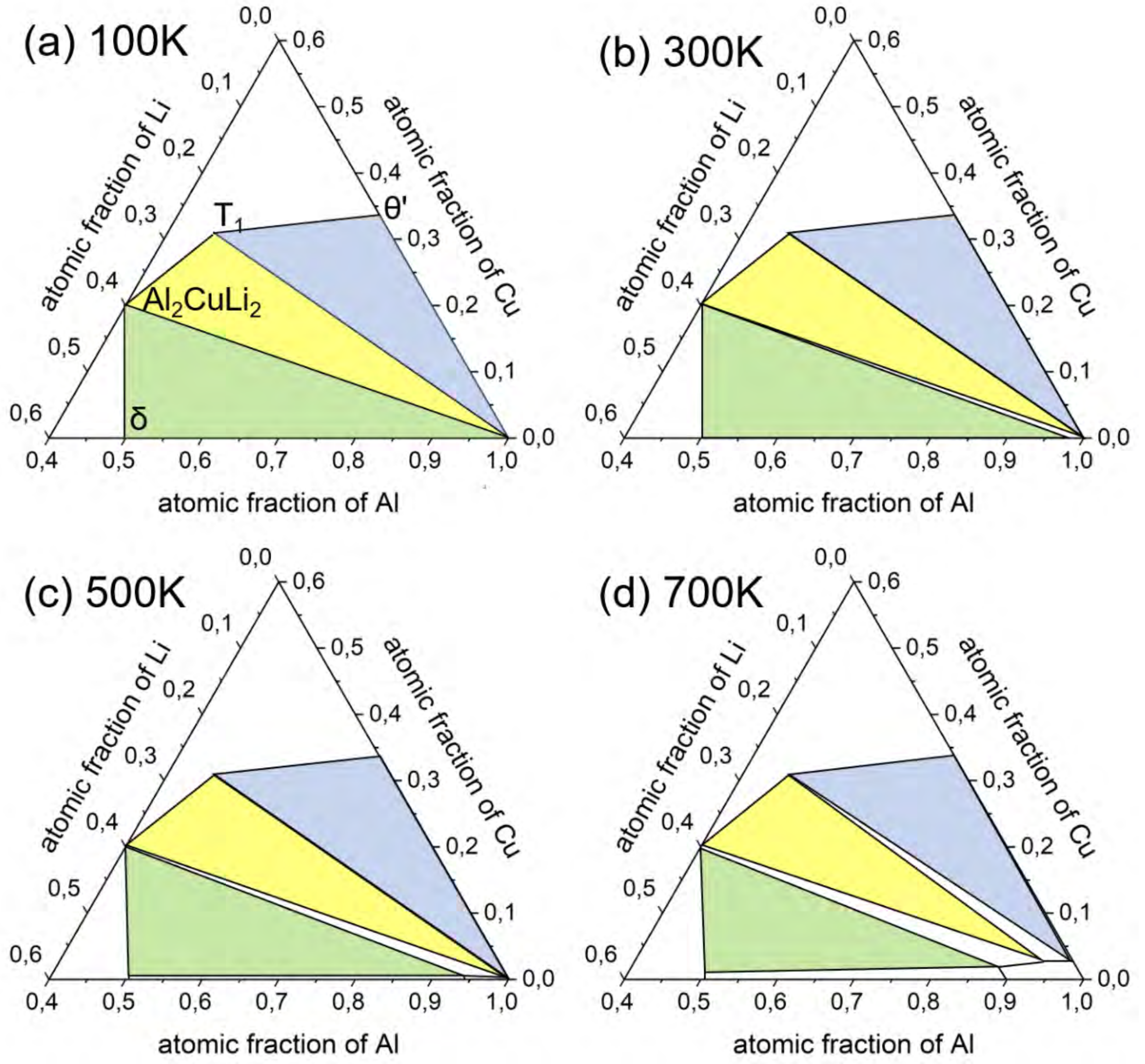

**Fig. 11.** Isothermal sections of Al-Li-Cu ternary phase diagram at different temperatures. The blue triangle stands for the equilibrium region between α-Al, θ' ($Al_2Cu$) and $T_1$. The yellow triangle stands for the equilibrium region between α-Al, $T_1$ and $Al_2CuLi_2$. The green triangle stands for the equilibrium region between α-Al, $Al_2CuLi_2$ and δ (AlLi).

The phase δ' ($Al_3Li$), which was a stable phase at 0 K (Fig. 10(d)), is not stable at higher temperatures. The distance from the formation enthalpy of δ' to the straight line connecting the formation enthalpies of α-Al and δ (AlLi) at 0 K was only -6.5 meV/atom, which is of the same order of the accuracy of DFT calculations. This distance changed to 12 meV/atom at 100 K, indicating that it becomes metastable at very low temperature. Therefore, δ' ($Al_3Li$) may not be a ground state phase, as indicated in previous investigations [60].

The line connecting Al and $Al_2CuLi_2$ changes to a very thin triangle at 200K (Fig. S2(a)), and the area of this triangle increases with temperature. This indicates that Li atoms dissolve into the Al matrix and the solubility increases with temperature, as reflected by the widening of the $Al_2CuLi_2$/α-Al two-phase region. Cu atoms start to be dissolved in the α-Al matrix at temperatures >500 K (Fig. 11(c)), leading to the apparition of a white polygonal zone near the Al matrix corner formed by a solid solution of Cu and Li into the α-Al matrix. The solubility of Cu also increases with temperature. It should be noted, however, that the solubility of Li in α-Al increases much faster with temperature than that of Cu.

The two-phase region represented by the line connecting δ (AlLi) and α-Al is transformed into a polygon at 500K (Fig. 11(c)), which indicates that Cu atoms are dissolved into δ (AlLi). However, the solubility of Cu in δ (AlLi) is very small even at very high temperature (Fig. 11(d)). Finally, the points associated with the phases θ' ($Al_2Cu$) and $T_1$ ($Al_6Cu_4Li_3$) do not show any changes in the isothermals at different temperatures, indicating that both phases are line compounds.

## 4. Discussion

### 4.1 Comparison with experimental phase diagrams

The experimental information about the Al-Li-Cu phase diagram is limited because of the difficulties associated with the large area to be explored in terms of different chemical compositions as well as by the limitations imposed by slow kinetics at low temperatures and the presence of metastable phases. Nevertheless, our results can be compared with the most recent experimental phase diagram for the Al-Li-Cu system summarized by Zinkevich *et al.* [6]. They reported that the solubility of Li and Cu in α-Al are 8 at. % and 1 at. % at 672 K. At 772 K, they reach 12 at. % and 2.5 at.% respectively. The solubilities of Li and Cu in our calculated phase diagram are 10 at.% (Li) and 3 at.% (Cu) at 700 K and 12 at.% (Li) and 3.5 at.% (Cu) at 800K, which are very similar to the experimental ones.

The experimental data in [6] indicate that the off-stoichiometry of δ (AlLi) reaches $Al_{0.54}Li_{0.46}$ at 673K and is practically the same at 773K. These results are very close to those in Fig. 11(d) and Fig. S2(d), which show that stoichiometry of δ is $Al_{0.52}Li_{0.48}$ at 700K and does not change at 800K. The $T_1$ phase in the experimental phase diagram is also treated as a line compound with an $Al_2CuLi$ stoichiometry and an (experimentally determined) enthalpy of formation of -20.4 kJ/mol (-0.21 eV/atom) [65,66]. This enthalpy of formation is very close to the formation enthalpy calculated by DFT for the $T_1$ phase, although the stoichiometry of this phase according to our theoretical calculations is $Al_6Cu_4Li_3$. It should be noted that the accurate experimental determination of the stoichiometry of these complex phases is very difficult.

$T_2$ and $T_B$ phases are also treated as line compounds in the experimental phase diagram [6]. Our theoretical analysis indicates that $T_B$ ($Al_7Cu_4Li$) configurations are not ground states, and the one with the lowest energy (Fig. 7(b)) is above the convex hull of the $CaF_2$ lattice by 0.06 eV/atom. Because the adjacent $Al_3Cu_2Li$ on the $CaF_2$ convex hull is not a global ground state phase, the distance of the $Al_7Cu_4Li$ structure in Fig. 7(b) to the global convex hull is even larger, 0.077 eV/atom. Thus, the formation enthalpy of $T_B$ ($Al_7Cu_4Li$) should be as low as -0.210 eV/atom to be on the global convex hull and become stable or it should be as low as -0.195 eV/atom to appear on the $CaF_2$ convex hull and become metastable. The formation enthalpy of very large cells with the $CaF_2$ lattice and $Al_7Cu_4Li$ stoichiometry (up to 96 atoms) was explored using the CE formalism but no one has a formation enthalpy lower than that in Fig. 7(b). Moreover, the experimentally determined enthalpy of formation for $T_B$ was -17 kJ/mol (-0.18 eV/atom) [6], which is close to the metastability condition but far away from the stability one. Thus, it is reasonable to assume that $T_B$ is metastable [43] and it should not be considered as stable phase in the phase diagram. However, the actual structure of $T_B$ should be explored further using very large cells.

We should notice that the structure of the $T_2$ phase in Fig. 2(i) contains 160 atoms. In this unit cell, there are 148 atomic sites that can be occupied by two different species, leading to a huge number of different configurations with the structure of the $T_2$ phase and the $Al_6CuLi_3$ stoichiometry. Thus, the exploration of the minimum energy configuration will require an unattainable number of DFT calculations and much more simulations will be necessary to develop a CE model. Therefore, the thermodynamic stability of $T_2$ phase was not assessed in this paper. Nevertheless, the experimentally determined enthalpy of formation $T_2$ is -20.4 kJ/mol (-0.21 eV/atom) [66], which indicates that it should be a stable phase. It should also be noticed that the $T_2$ phase - as other quasicrystals with icosahedral symmetry in Al alloys - develops from the melt [36, 41, 67] but it is difficult to nucleate from thermo-mechanical treatments due to the large size of the unit cell. So, its influence of the strengthening mechanisms is limited.

The first-principles calculations also indicate that the phases with simple crystallographic lattices, such as fcc (α-Al) and bcc (δ and $Al_2CuLi_2$), are off-stoichiometry at elevated temperature (Fig. 11 and Fig. S2). However, θ' ($Al_2Cu$) and $T_1$ ($Al_6Cu_4Li_3$), that present more complex lattice structures ($CaF_2$ and hexagonal, respectively), are line compounds. Line compounds are characterized by a considerable lower formation enthalpy with respect to other configurations with the same composition [68] but this is not the case for θ' ($Al_2Cu$) and $T_1$ ($Al_6Cu_4Li_3$). Following the theoretical analysis of Krivovichev [69], the configurational entropy contribution to the free energy of complicated crystal structures is smaller than that of simpler crystal structures, and this is the reason why θ' ($Al_2Cu$) and $T_1$ ($Al_6Cu_4Li_3$) are line compounds. Therefore, $T_2$ and $T_B$ are predicted to be line compounds because of their more complicated structures, in agreement with the experimental phase diagram of Zinkevich et al. [6].

Our calculated phase diagram differs from the experimental one [6] in the stability of θ ($Al_2Cu$) and θ' ($Al_2Cu$). The formation enthalpy of θ' ($Al_2Cu$) is 0.012 eV/atom lower than that of θ ($Al_2Cu$) at 0K (Fig. 10(a)) according to the first-principles calculations, in agreement with previous results [11-14]. If only the configurational entropy contribution is considered, θ' ($Al_2Cu$) remains the stable phase at high temperatures, but this result is not in agreement with the experimental evidence in binary Al-Cu alloys. In fact, the small difference in formation enthalpies between θ' and θ at high temperature is reversed because of the vibrational entropic contribution at approximately 550K [11, 13-14] but this effect was not considered in our theoretical model. From current work in Al-Li-Cu ternary system, the stability from configurational entropy contribution is the same as that in Al-Cu binary system, indicating the contribution of Li to the stability θ ($Al_2Cu$) and θ' ($Al_2Cu$) is negligible.

### 4.2 Effects on precipitation

The formation enthalpies as well as the calculated phase diagram can be used to rationalize the precipitation mechanisms in Al-Li-Cu alloys during high temperature aging. According to the global convex hull of Al-Li-Cu system, θ'' ($Al_3Cu$) is not a ground state phase and its formation enthalpy is 0.036 eV/atom above the line connecting Al and θ' ($Al_2Cu$). This magnitude is not negligible, taking into the lowest formation enthalpies calculated for Al-Li-Cu system (-0.23 eV/atom). However, the precipitation of θ'' ($Al_3Cu$) is frequently observed in Al-Li-Cu alloys during low temperature aging [70], regardless of the larger chemical energy barrier for precipitate nucleation than θ' ($Al_2Cu$). This phenomenon can be rationalized because θ'' ($Al_3Cu$) shares the same fcc lattice (with similar lattice parameters) and its coherent with α-Al. Thus, the interface and elastic mismatch energies for the nucleation of θ'' in α-Al are very small (as compared with those of θ' and θ) and they favor homogeneous nucleation. On the contrary, the homogeneous nucleation of θ' is very difficult because of the energy associated with the elastic strains induced in the matrix, even though is more stable from the thermodynamic viewpoint that θ'', and θ' and θ tend to nucleate heterogeneously at grain boundaries, dislocations and precipitate after the homogeneous nucleation of θ'' [12]. A similar argument can be used to explain why homogeneous precipitation of δ' ($Al_3Li$) with fcc structure, which is a slightly metastable phase, occurs sooner than the heterogenous precipitation of δ (AlLi) on δ' precipitates. We should note here the common feature of θ'' and δ' is they are located on the local convex hull of each lattice structure. This is why they are denominated ‘phase’ instead of transition state (as Guinier-Preston zones).

Another reason for the early precipitation of θ'' ($Al_3Cu$) and δ' ($Al_3Li$) in the Al-Li-Cu ternary system can be found because of their location in the ternary composition space. Regardless of whether they are in the local or global convex hull, both are the closest phases to the Al corner. So, nucleation of these phases may occur without overcoming other energy barriers. A similar argument can be used to explain the precipitation of the metastable $T_1'$ ($Al_2CuLi$) phase. Although this phase is above the global convex hull, it is on the local $CaF_2$ convex hull and also has a tie-line with α-Al. Therefore, the transition from α-Al to $T_1'$ ($Al_2CuLi$) does not need to cross other energy barriers. Moreover, the formation enthalpy of $T_1'$ ($Al_2CuLi$) is very close to the line connecting the formation enthalpies of the stable $T_1$ ($Al_6Cu_4Li_3$) and α-Al phases, and $T_1'$ ($Al_2CuLi$) has been identified as the precursor precipitate of $T_1$ [32], as θ'' and δ' precede the precipitation of θ' and δ, respectively [10, 15].

After aging at higher temperature or for longer times, the stable precipitate phases in the Al-Li-Cu system are θ' ($Al_2Cu$), δ(AlLi) and $T_1$ ($Al_6Cu_4Li_3$) according to the phase diagram. Thus, θ' and δ precipitates always appear. Nevertheless, it should be noticed that previous calculations of the free energy including the vibrational entropy contribution in the binary Al-Cu system showed that stable phase with the $Al_2Cu$ stoichiometry changes from θ' to θ at 550 K [11, 13]. This is possible because the differences in the configurational free energy between both phases are very small (always below 15 meV/atom) and can be overcome by the vibrational entropic contribution at high temperature. Precipitation of $T_1$ ($Al_6Cu_4Li_3$), however, is hindered because of the large size of the unit cell that requires long range ordering even though it is a stable phase and has a tie-line connection with α-Al in the phase diagram (Fig. 11 and Fig. S2). Actually, a number of investigations [71-75] reported that $T_1$ ($Al_6Cu_4Li_3$) can only be obtained by rapid or slow cooling from the melt, and dislocations need to be introduced by pre-deformation to help the nucleation of $T_1$ during solid phase transformations [76]. Similarly, the $T_B$ ($Al_7Cu_4Li$) phase is metastable and has a large unit cell and its precipitation will also be hindered even though it has a tie-line connection to α-Al (Fig. 1).

## 5. Conclusions

The phase diagram of Al-Li-Cu system in the Al-rich region was determined by means of first-principles calculations. To this end, the mixing enthalpies of different configurations for various lattices in the whole Al-Li-Cu system were determined by density functional theory simulations which indicated the stable phases in the convex hull. In addition, the mixing enthalpies were fitted with a cluster expansion that was used to calculate the free energy of the lattice with different compositions as a function of temperature in the Al-rich region (Al content > 40 at. %) by means of Monte Carlo simulations and statistical mechanics. The common tangent to the lowest free energy surfaces at each temperature was used to determine the phase boundaries.

In general, the calculated phase diagram was in good agreement with the limited experimental information. In particular, it was found that the ground state phases in the Al-rich part of the Al-Li-Cu phase diagram were α-Al, θ' ($Al_2Cu$), δ' ($Al_3Li$), δ (AlLi) and $T_1$ ($Al_6Cu_4Li_3$), while θ'' ($Al_3Cu$), $T_1'$ ($Al_2CuLi$) and $Al_3Cu_2Li$ were found on the lowest mixing enthalpy surfaces of their lattices and were metastable. α-Al, δ and $T_1$ are stable phases in the whole temperature range while δ' becomes metastable at very low temperature and θ ($Al_2Cu$) replaces θ' as the stable phase with this stoichiometry at approximately 550K due to the vibrational entropic contribution. The assessment of the thermodynamic stability of $T_2$ ($Al_6CuLi_3$) is beyond the capabilities of first-principles simulations because of the large size of the unit cell (160 atoms) but it has been experimentally reported as a stable phase. It should be noted that the first-principles simulations predict that the stoichiometry of $T_1$ is $Al_6Cu_4Li_3$ and not $Al_2CuLi$, as reported experimentally. Moreover, our calculations showed that the structures of $T_1'$ ($Al_2CuLi$) and $T_B$ ($Al_7Cu_4Li$) are bcc and $CaF_2$, respectively.

The isothermal sections of the phase diagram of the Al-Li-Cu system in the Al-rich region from 100K to 900K included three triangular regions which stand for the stability region for the three different phases at the corners of each triangle. They are α-Al, θ' and $T_1$; α-Al, $T_1$ and $Al_2CuLi_2$; and α-Al, $Al_2CuLi_2$ and δ (AlLi). In addition, lines or new triangular zones (that appear at high temperature) between each two three-phase equilibrium regions are the two-phase regions, while a polygonal region also appears near the Al-rich corner at T > 500 K as a result of the solid solution of Li and Cu in the α-Al matrix. The first-principles simulations indicate that the phases with simple crystallographic lattices, such as fcc (α-Al) and bcc (δ and $Al_2CuLi_2$), are off-stoichiometry at elevated temperature, in agreement with experimental results. On the contrary, the phases with a more complex lattice - such as $T_1$ (hexagonal), θ' (CaF2), $T_2$ and $T_B$- are predicted to be line compounds.

## 6. Acknowledgements

This investigation was supported by the European Union's Horizon 2020 research and innovation programme through a Marie Sklodowska-Curie Individual Fellowship (Grant Agreement 893883). Computer resources and technical assistance provided by the Centro de Supercomputación y Visualización de Madrid (CeSViMa) and by the Spanish Supercomputing Network (project FI-2021-3-6) are gratefully acknowledged. Finally, use of the computational resources of the Center for Nanoscale Materials, an Office of Science user facility, supported by the U.S. Department of Energy, Office of Science, Office of Basic Energy Sciences, under Contract No. DE-AC02-06CH11357, is also gratefully acknowledged.

**SUPPLEMENTARY INFORMATION**

**First principles analysis of the Al-rich corner of Al-Li-Cu phase diagram**

S. Liu, J. S. Wróbel, J. LLorca

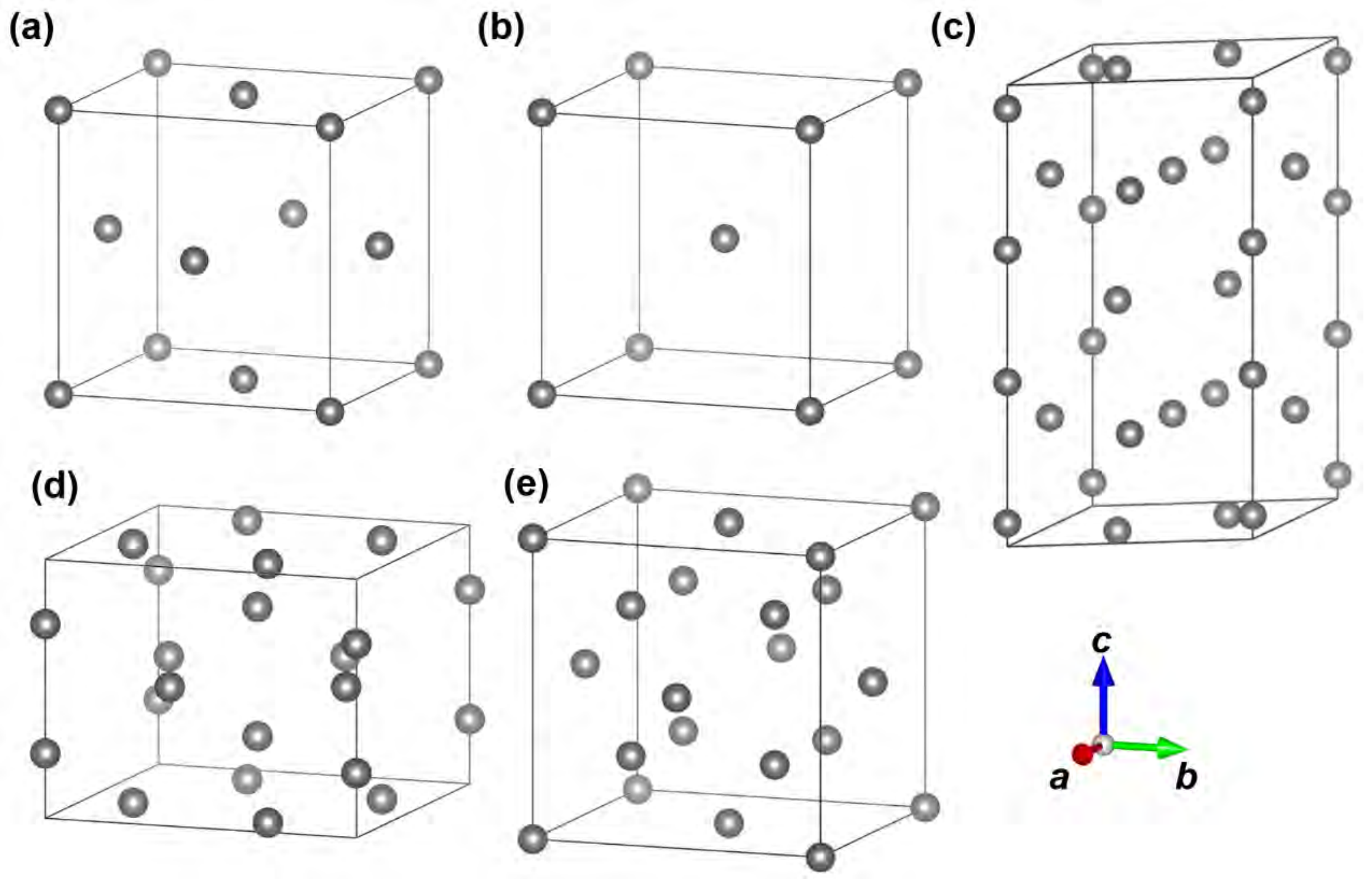

**Fig. S1.** Crystal lattices of the phases in the Al-rich corner of the Al-Li-Cu system.
(a) fcc. (b) bcc. (c) hexagonal. (d) bct. (e) cubic $CaF_2$.

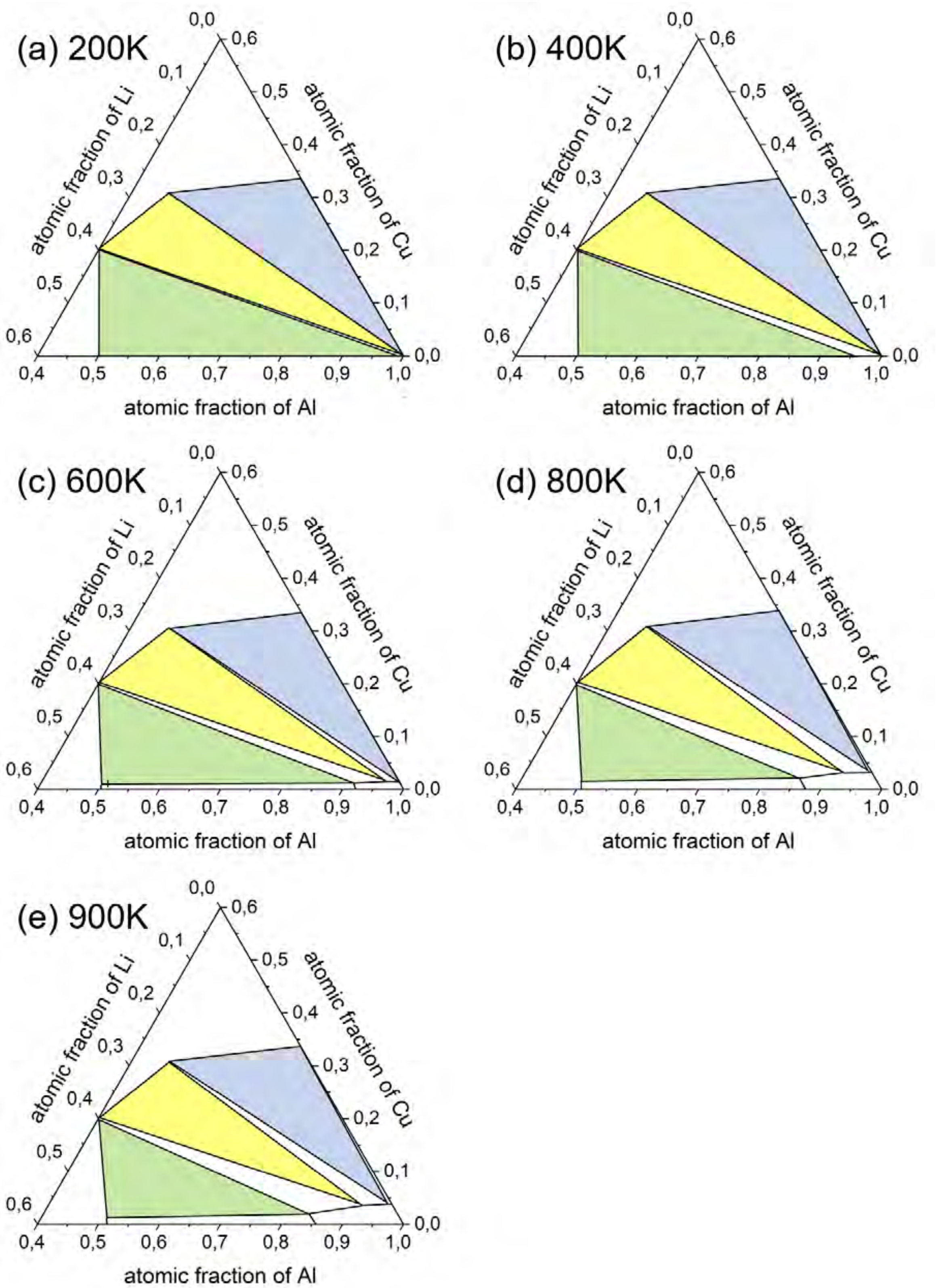

**Fig. S2.** Isothermal sections of Al-Li-Cu ternary phase diagram at different temperatures. The blue triangle stands for the equilibrium region between α-Al, θ' ($Al_2Cu$) and $T_1$. The yellow triangle stands for the equilibrium region between α-Al, $T_1$ and $Al_2CuLi_2$. The green triangle stands for the equilibrium region between α-Al, $Al_2CuLi_2$ and δ (AlLi).

**Table S1.** Cluster expansion for fcc lattice.

| cluster | labels | multiplicity | coordinates | point function index | ECI |
|---|---|---|---|---|---|
| empty | 1 | 1 | | | -0.117871 |
| point | 1 | 1 | 2, 2, 2 | 0 | 0.029694 |
| | 2 | 1 | 2, 2, 2 | 1 | 0.042897 |
| pair | 1 | 6 | 2, 2, 2 | 0 | -0.006659 |
| | | | 1, 2, 3 | 0 | |
| | 2 | 12 | 2, 2, 2 | 1 | 0.005090 |
| | | | 1, 2, 3 | 0 | |
| | 3 | 6 | 2, 2, 2 | 1 | 0.008765 |
| | | | 1, 2, 3 | 1 | |
| | 4 | 3 | 2, 2, 2 | 0 | 0.004273 |
| | | | 0, 2, 2 | 0 | |
| | 5 | 6 | 2, 2, 2 | 1 | -0.007263 |
| | | | 0, 2, 2 | 0 | |
| | 6 | 3 | 2, 2, 2 | 1 | 0.006333 |
| | | | 0, 2, 2 | 1 | |
| | 7 | 12 | 2, 2, 2 | 0 | -0.000279 |
| | | | 0, 3, 3 | 0 | |
| | 8 | 24 | 2, 2, 2 | 1 | -0.001442 |
| | | | 0, 3, 3 | 0 | |
| | 9 | 12 | 2, 2, 2 | 1 | 0.002120 |
| | | | 0, 3, 3 | 1 | |
| | 10 | 6 | 2, 2, 2 | 0 | -0.000355 |
| | | | 0, 2, 4 | 0 | |
| | 11 | 12 | 2, 2, 2 | 1 | 0.005868 |
| | | | 0, 2, 4 | 0 | |
| | 12 | 6 | 2, 2, 2 | 1 | -0.001212 |
| | | | 0, 2, 4 | 1 | |
| | 13 | 12 | 2, 2, 2 | 0 | 0.001760 |
| | | | -1, 2, 3 | 0 | |
| | 14 | 24 | 2, 2, 2 | 1 | -0.001074 |
| | | | -1, 2, 3 | 0 | |
| | 15 | 12 | 2, 2, 2 | 1 | -0.001201 |
| | | | -1, 2, 3 | 1 | |
| | 16 | 4 | 2, 2, 2 | 0 | 0.002387 |
| | | | 0, 4, 4 | 0 | |
| | 17 | 8 | 2, 2, 2 | 1 | -0.001375 |
| | | | 0, 4, 4 | 0 | |
| | 18 | 4 | 2, 2, 2 | 1 | -0.000713 |
| | | | 0, 4, 4 | 1 | |
| | 19 | 24 | 2, 2, 2 | 0 | 0.000257 |
| | | | -1, 3, 4 | 0 | |
| | 20 | 48 | 2, 2, 2 | 1 | -0.000057 |
| | | | -1, 3, 4 | 0 | |
| | 21 | 24 | 2, 2, 2 | 1 | -0.000926 |
| | | | -1, 3, 4 | 1 | |
| | 22 | 3 | 2, 2, 2 | 0 | 0.001930 |
| | | | 2, 2, 6 | 0 | |
| | 23 | 6 | 2, 2, 2 | 1 | -0.002955 |
| | | | 2, 2, 6 | 0 | |
| | 24 | 3 | 2, 2, 2 | 1 | 0.009750 |
| | | | 2, 2, 6 | 1 | |
| triplet | 1 | 8 | 2, 2, 2 | 0 | -0.005199 |
| | | | 1, 2, 3 | 0 | |
| | | | 2, 3, 3 | 0 | |
| | 2 | 24 | 2, 2, 2 | 1 | 0.001282 |
| | | | 1, 2, 3 | 0 | |
| | | | 2, 3, 3 | 0 | |
| | 3 | 24 | 2, 2, 2 | 1 | -0.004435 |
| | | | 1, 2, 3 | 1 | |
| | | | 2, 3, 3 | 0 | |
| | 4 | 8 | 2, 2, 2 | 1 | 0.006998 |
| | | | 1, 2, 3 | 1 | |
| | | | 2, 3, 3 | 1 | |
| | 5 | 12 | 2, 2, 2 | 0 | 0.001048 |
| | | | 1, 2, 3 | 0 | |
| | | | 0, 2, 2 | 0 | |
| | 6 | 24 | 2, 2, 2 | 1 | -0.003547 |
| | | | 1, 2, 3 | 0 | |
| | | | 0, 2, 2 | 0 | |
| | 7 | 12 | 2, 2, 2 | 0 | -0.000767 |
| | | | 1, 2, 3 | 1 | |
| | | | 0, 2, 2 | 0 | |
| | 8 | 24 | 2, 2, 2 | 1 | 0.001707 |
| | | | 1, 2, 3 | 1 | |
| | | | 0, 2, 2 | 0 | |
| | 9 | 12 | 2, 2, 2 | 1 | 0.000898 |
| | | | 1, 2, 3 | 0 | |
| | | | 0, 2, 2 | 1 | |
| | 10 | 12 | 2, 2, 2 | 1 | 0.000018 |
| | | | 1, 2, 3 | 1 | |
| | | | 0, 2, 2 | 1 | |
| | 11 | 24 | 2, 2, 2 | 0 | -0.001878 |
| | | | 2, 3, 3 | 0 | |
| | | | 0, 3, 3 | 0 | |
| | 12 | 24 | 2, 2, 2 | 1 | 0.003843 |
| | | | 2, 3, 3 | 0 | |
| | | | 0, 3, 3 | 0 | |
| | 13 | 24 | 2, 2, 2 | 0 | -0.001218 |
| | | | 2, 3, 3 | 1 | |
| | | | 0, 3, 3 | 0 | |
| | 14 | 24 | 2, 2, 2 | 1 | -0.003094 |
| | | | 2, 3, 3 | 1 | |
| | | | 0, 3, 3 | 0 | |
| | 15 | 24 | 2, 2, 2 | 0 | 0.001909 |
| | | | 2, 3, 3 | 0 | |
| | | | 0, 3, 3 | 1 | |
| | 16 | 24 | 2, 2, 2 | 1 | 0.001605 |
| | | | 2, 3, 3 | 0 | |
| | | | 0, 3, 3 | 1 | |
| | 17 | 24 | 2, 2, 2 | 0 | 0.000266 |
| | | | 2, 3, 3 | 1 | |
| | | | 0, 3, 3 | 1 | |
| | 18 | 24 | 2, 2, 2 | 1 | -0.000928 |
| | | | 2, 3, 3 | 1 | |
| | | | 0, 3, 3 | 1 | |
| | 19 | 24 | 2, 2, 2 | 0 | 0.001162 |
| | | | 1, 2, 3 | 0 | |
| | | | 0, 3, 3 | 0 | |
| | 20 | 48 | 2, 2, 2 | 1 | -0.004183 |
| | | | 1, 2, 3 | 0 | |
| | | | 0, 3, 3 | 0 | |
| | 21 | 24 | 2, 2, 2 | 0 | 0.001060 |
| | | | 1, 2, 3 | 1 | |
| | | | 0, 3, 3 | 0 | |
| | 22 | 48 | 2, 2, 2 | 1 | -0.001111 |
| | | | 1, 2, 3 | 1 | |
| | | | 0, 3, 3 | 0 | |
| | 23 | 24 | 2, 2, 2 | 1 | 0.000550 |
| | | | 1, 2, 3 | 0 | |
| | | | 0, 3, 3 | 1 | |
| | 24 | 24 | 2, 2, 2 | 1 | 0.001263 |
| | | | 1, 2, 3 | 1 | |
| | | | 0, 3, 3 | 1 | |
| | 25 | 24 | 2, 2, 2 | 0 | 0.000537 |
| | | | 1, 1, 2 | 0 | |
| | | | 0, 3, 3 | 0 | |
| | 26 | 48 | 2, 2, 2 | 1 | -0.001491 |
| | | | 1, 1, 2 | 0 | |
| | | | 0, 3, 3 | 0 | |
| | 27 | 24 | 2, 2, 2 | 1 | -0.000471 |
| | | | 1, 1, 2 | 1 | |
| | | | 0, 3, 3 | 0 | |
| | 28 | 24 | 2, 2, 2 | 0 | -0.001538 |
| | | | 1, 1, 2 | 0 | |
| | | | 0, 3, 3 | 1 | |
| | 29 | 48 | 2, 2, 2 | 1 | -0.000424 |
| | | | 1, 1, 2 | 0 | |
| | | | 0, 3, 3 | 1 | |
| | 30 | 24 | 2, 2, 2 | 1 | -0.001004 |
| | | | 1, 1, 2 | 1 | |
| | | | 0, 3, 3 | 1 | |
| | 31 | 24 | 2, 2, 2 | 0 | 0.000364 |
| | | | 2, 2, 4 | 0 | |
| | | | 0, 3, 3 | 0 | |
| | 32 | 48 | 2, 2, 2 | 1 | -0.000057 |
| | | | 2, 2, 4 | 0 | |
| | | | 0, 3, 3 | 0 | |
| | 33 | 24 | 2, 2, 2 | 1 | 0.000124 |
| | | | 2, 2, 4 | 1 | |
| | | | 0, 3, 3 | 0 | |
| | 34 | 24 | 2, 2, 2 | 0 | 0.000367 |
| | | | 2, 2, 4 | 0 | |
| | | | 0, 3, 3 | 1 | |
| | 35 | 48 | 2, 2, 2 | 1 | -0.000999 |
| | | | 2, 2, 4 | 0 | |
| | | | 0, 3, 3 | 1 | |
| | 36 | 24 | 2, 2, 2 | 1 | 0.000703 |
| | | | 2, 2, 4 | 1 | |
| | | | 0, 3, 3 | 1 | |
| | 37 | 8 | 2, 2, 2 | 0 | 0.000426 |
| | | | 1, 4, 1 | 0 | |
| | | | 0, 3, 3 | 0 | |
| | 38 | 24 | 2, 2, 2 | 1 | 0.000493 |
| | | | 1, 4, 1 | 0 | |
| | | | 0, 3, 3 | 0 | |
| | 39 | 24 | 2, 2, 2 | 1 | 0.000284 |
| | | | 1, 4, 1 | 1 | |
| | | | 0, 3, 3 | 0 | |
| | 40 | 8 | 2, 2, 2 | 1 | 0.000938 |
| | | | 1, 4, 1 | 1 | |
| | | | 0, 3, 3 | 1 | |
| | 41 | 48 | 2, 2, 2 | 0 | -0.001133 |
| | | | 2, 3, 3 | 0 | |
| | | | 0, 2, 4 | 0 | |
| | 42 | 48 | 2, 2, 2 | 1 | 0.001749 |
| | | | 2, 3, 3 | 0 | |
| | | | 0, 2, 4 | 0 | |
| | 43 | 48 | 2, 2, 2 | 0 | 0.000489 |
| | | | 2, 3, 3 | 1 | |
| | | | 0, 2, 4 | 0 | |
| | 44 | 48 | 2, 2, 2 | 1 | -0.000941 |
| | | | 2, 3, 3 | 1 | |
| | | | 0, 2, 4 | 0 | |
| | 45 | 48 | 2, 2, 2 | 0 | 0.000145 |
| | | | 2, 3, 3 | 0 | |
| | | | 0, 2, 4 | 1 | |
| | 46 | 48 | 2, 2, 2 | 1 | 0.000153 |
| | | | 2, 3, 3 | 0 | |
| | | | 0, 2, 4 | 1 | |
| | 47 | 48 | 2, 2, 2 | 0 | 0.000208 |
| | | | 2, 3, 3 | 1 | |
| | | | 0, 2, 4 | 1 | |
| | 48 | 48 | 2, 2, 2 | 1 | -0.005842 |
| | | | 2, 3, 3 | 1 | |
| | | | 0, 2, 4 | 1 | |
| | 49 | 6 | 2, 2, 2 | 0 | 0.004116 |
| | | | 1, 2, 3 | 0 | |
| | | | 0, 2, 4 | 0 | |
| | 50 | 12 | 2, 2, 2 | 1 | 0.000507 |
| | | | 1, 2, 3 | 0 | |
| | | | 0, 2, 4 | 0 | |
| | 51 | 6 | 2, 2, 2 | 0 | -0.000068 |
| | | | 1, 2, 3 | 1 | |
| | | | 0, 2, 4 | 0 | |
| | 52 | 12 | 2, 2, 2 | 1 | -0.002566 |
| | | | 1, 2, 3 | 1 | |
| | | | 0, 2, 4 | 0 | |
| | 53 | 6 | 2, 2, 2 | 1 | -0.000151 |
| | | | 1, 2, 3 | 0 | |
| | | | 0, 2, 4 | 1 | |
| | 54 | 6 | 2, 2, 2 | 1 | 0.000638 |
| | | | 1, 2, 3 | 1 | |
| | | | 0, 2, 4 | 1 | |
| | 55 | 12 | 2, 2, 2 | 0 | 0.001155 |
| | | | 0, 2, 2 | 0 | |
| | | | 0, 2, 4 | 0 | |
| | 56 | 24 | 2, 2, 2 | 1 | -0.004276 |
| | | | 0, 2, 2 | 0 | |
| | | | 0, 2, 4 | 0 | |
| | 57 | 12 | 2, 2, 2 | 0 | 0.000735 |
| | | | 0, 2, 2 | 1 | |
| | | | 0, 2, 4 | 0 | |
| | 58 | 24 | 2, 2, 2 | 1 | 0.000352 |
| | | | 0, 2, 2 | 1 | |
| | | | 0, 2, 4 | 0 | |
| | 59 | 12 | 2, 2, 2 | 1 | 0.000093 |
| | | | 0, 2, 2 | 0 | |
| | | | 0, 2, 4 | 1 | |
| | 60 | 12 | 2, 2, 2 | 1 | 0.000793 |
| | | | 0, 2, 2 | 1 | |
| | | | 0, 2, 4 | 1 | |
| | 61 | 12 | 2, 2, 2 | 0 | -0.000241 |
| | | | 1, 4, 3 | 0 | |
| | | | 0, 2, 4 | 0 | |
| | 62 | 24 | 2, 2, 2 | 1 | -0.001834 |
| | | | 1, 4, 3 | 0 | |
| | | | 0, 2, 4 | 0 | |
| | 63 | 12 | 2, 2, 2 | 0 | 0.001411 |
| | | | 1, 4, 3 | 1 | |
| | | | 0, 2, 4 | 0 | |
| | 64 | 24 | 2, 2, 2 | 1 | 0.000436 |
| | | | 1, 4, 3 | 1 | |
| | | | 0, 2, 4 | 0 | |
| | 65 | 12 | 2, 2, 2 | 1 | 0.000263 |
| | | | 1, 4, 3 | 0 | |
| | | | 0, 2, 4 | 1 | |
| | 66 | 12 | 2, 2, 2 | 1 | 0.000093 |
| | | | 1, 4, 3 | 1 | |
| | | | 0, 2, 4 | 1 | |
| | 67 | 8 | 2, 2, 2 | 0 | 0.000100 |
| | | | 0, 4, 2 | 0 | |
| | | | 0, 2, 4 | 0 | |
| | 68 | 24 | 2, 2, 2 | 1 | -0.000650 |
| | | | 0, 4, 2 | 0 | |
| | | | 0, 2, 4 | 0 | |
| | 69 | 24 | 2, 2, 2 | 1 | 0.000173 |
| | | | 0, 4, 2 | 1 | |
| | | | 0, 2, 4 | 0 | |
| | 70 | 8 | 2, 2, 2 | 1 | 0.000230 |
| | | | 0, 4, 2 | 1 | |
| | | | 0, 2, 4 | 1 | |
| | 71 | 24 | 2, 2, 2 | 0 | -0.001078 |
| | | | 2, 3, 3 | 0 | |
| | | | -1, 2, 3 | 0 | |
| | 72 | 48 | 2, 2, 2 | 1 | -0.000012 |
| | | | 2, 3, 3 | 0 | |
| | | | -1, 2, 3 | 0 | |
| | 73 | 24 | 2, 2, 2 | 1 | -0.000267 |
| | | | 2, 3, 3 | 1 | |
| | | | -1, 2, 3 | 0 | |
| | 74 | 24 | 2, 2, 2 | 0 | 0.000465 |
| | | | 2, 3, 3 | 0 | |
| | | | -1, 2, 3 | 1 | |
| | 75 | 48 | 2, 2, 2 | 1 | -0.000267 |
| | | | 2, 3, 3 | 0 | |
| | | | -1, 2, 3 | 1 | |
| | 76 | 24 | 2, 2, 2 | 1 | -0.000644 |
| | | | 2, 3, 3 | 1 | |
| | | | -1, 2, 3 | 1 | |
| | 77 | 24 | 2, 2, 2 | 0 | -0.000237 |
| | | | 1, 2, 3 | 0 | |
| | | | -1, 2, 3 | 0 | |
| | 78 | 24 | 2, 2, 2 | 1 | 0.000941 |
| | | | 1, 2, 3 | 0 | |
| | | | -1, 2, 3 | 0 | |
| | 79 | 24 | 2, 2, 2 | 0 | 0.001308 |
| | | | 1, 2, 3 | 1 | |
| | | | -1, 2, 3 | 0 | |
| | 80 | 24 | 2, 2, 2 | 1 | -0.000518 |
| | | | 1, 2, 3 | 1 | |
| | | | -1, 2, 3 | 0 | |
| | 81 | 24 | 2, 2, 2 | 0 | 0.002329 |
| | | | 1, 2, 3 | 0 | |
| | | | -1, 2, 3 | 1 | |
| | 82 | 24 | 2, 2, 2 | 1 | -0.000306 |
| | | | 1, 2, 3 | 0 | |
| | | | -1, 2, 3 | 1 | |
| | 83 | 24 | 2, 2, 2 | 0 | -0.000378 |
| | | | 1, 2, 3 | 1 | |
| | | | -1, 2, 3 | 1 | |
| | 84 | 24 | 2, 2, 2 | 1 | -0.000899 |
| | | | 1, 2, 3 | 1 | |
| | | | -1, 2, 3 | 1 | |

**Table S2.** Cluster expansion for bcc lattice.

| cluster | labels | multiplicity | coordinates | point function index | ECI |
|---|---|---|---|---|---|
| empty | 1 | 1 | | | -0.155308 |
| point | 1 | 1 | 3, 1, 1 | 0 | 0.068235 |
| | 2 | 1 | 3, 1, 1 | 1 | 0.071081 |
| pair | 1 | 4 | 3, 1, 1 | 0 | 0.091832 |
| | | | 4, 0, 2 | 0 | |
| | 2 | 8 | 3, 1, 1 | 1 | -0.019309 |
| | | | 4, 0, 2 | 0 | |
| | 3 | 4 | 3, 1, 1 | 1 | 0.001553 |
| | | | 4, 0, 2 | 1 | |
| | 4 | 3 | 3, 1, 1 | 0 | 0.024458 |
| | | | 1, 1, 1 | 0 | |
| | 5 | 6 | 3, 1, 1 | 1 | 0.012805 |
| | | | 1, 1, 1 | 0 | |
| | 6 | 3 | 3, 1, 1 | 1 | 0.016905 |
| | | | 1, 1, 1 | 1 | |
| | 7 | 6 | 3, 1, 1 | 0 | 0.010713 |
| | | | 1, -1, 1 | 0 | |
| | 8 | 12 | 3, 1, 1 | 1 | 0.005370 |
| | | | 1, -1, 1 | 0 | |
| | 9 | 6 | 3, 1, 1 | 1 | 0.005189 |
| | | | 1, -1, 1 | 1 | |
| | 10 | 12 | 3, 1, 1 | 0 | 0.003061 |
| | | | 0, 2, 2 | 0 | |
| | 11 | 24 | 3, 1, 1 | 1 | -0.001491 |
| | | | 0, 2, 2 | 0 | |
| | 12 | 12 | 3, 1, 1 | 1 | -0.002408 |
| | | | 0, 2, 2 | 1 | |
| | 13 | 4 | 3, 1, 1 | 0 | 0.006131 |
| | | | 1, 3, 3 | 0 | |
| | 14 | 8 | 3, 1, 1 | 1 | -0.001369 |
| | | | 1, 3, 3 | 0 | |
| | 15 | 4 | 3, 1, 1 | 1 | -0.006847 |
| | | | 1, 3, 3 | 1 | |
| | 16 | 3 | 3, 1, 1 | 0 | 0.008367 |
| | | | -1, 1, 1 | 0 | |
| | 17 | 6 | 3, 1, 1 | 1 | -0.005118 |
| | | | -1, 1, 1 | 0 | |
| | 18 | 3 | 3, 1, 1 | 1 | 0.002752 |
| | | | -1, 1, 1 | 1 | |
| | 19 | 12 | 3, 1, 1 | 0 | -0.006737 |
| | | | 0, 2, 4 | 0 | |
| | 20 | 24 | 3, 1, 1 | 1 | 0.005067 |
| | | | 0, 2, 4 | 0 | |
| | 21 | 12 | 3, 1, 1 | 1 | -0.000402 |
| | | | 0, 2, 4 | 1 | |
| | 22 | 12 | 3, 1, 1 | 0 | -0.001810 |
| | | | -1, 1, 3 | 0 | |
| | 23 | 24 | 3, 1, 1 | 1 | 0.001930 |
| | | | -1, 1, 3 | 0 | |
| | 24 | 12 | 3, 1, 1 | 1 | -0.001569 |
| | | | -1, 1, 3 | 1 | |
| triplet | 1 | 12 | 3, 1, 1 | 0 | 0.013941 |
| | | | 2, 2, 2 | 0 | |
| | | | 3, 1, 1 | 0 | |
| | 2 | 24 | 2, 2, 2 | 1 | -0.016711 |
| | | | 1, 1, 1 | 0 | |
| | | | 3, 1, 1 | 0 | |
| | 3 | 12 | 3, 1, 1 | 0 | 0.012231 |
| | | | 2, 2, 2 | 1 | |
| | | | 1, 1, 1 | 0 | |
| | 4 | 24 | 3, 1, 1 | 1 | 0.001474 |
| | | | 2, 2, 2 | 1 | |
| | | | 1, 1, 1 | 0 | |
| | 5 | 24 | 3, 1, 1 | 1 | -0.000487 |
| | | | 2, 2, 2 | 0 | |
| | | | 1, 1, 1 | 1 | |
| | 6 | 12 | 3, 1, 1 | 1 | -0.002468 |
| | | | 2, 2, 2 | 1 | |
| | | | 1, 1, 1 | 1 | |
| | 7 | 12 | 3, 1, 1 | 0 | -0.001131 |
| | | | 2, 0, 2 | 0 | |
| | | | 1, -1, 1 | 0 | |
| | 8 | 12 | 3, 1, 1 | 1 | 0.002502 |
| | | | 2, 0, 2 | 0 | |
| | | | 1, -1, 1 | 0 | |
| | 9 | 12 | 3, 1, 1 | 0 | 0.004286 |
| | | | 2, 0, 2 | 1 | |
| | | | 1, -1, 1 | 0 | |
| | 10 | 24 | 3, 1, 1 | 1 | 0.001256 |
| | | | 2, 0, 2 | 1 | |
| | | | 1, -1, 1 | 0 | |
| | 11 | 12 | 3, 1, 1 | 1 | -0.000580 |
| | | | 2, 0, 2 | 0 | |
| | | | 1, -1, 1 | 1 | |
| | 12 | 12 | 3, 1, 1 | 1 | -0.001106 |
| | | | 2, 0, 2 | 1 | |
| | | | 1, -1, 1 | 1 | |
| | 13 | 12 | 3, 1, 1 | 0 | -0.001371 |
| | | | 1, 1, 1 | 0 | |
| | | | 1, -1, 1 | 0 | |
| | 14 | 24 | 3, 1, 1 | 1 | -0.000422 |
| | | | 1, 1, 1 | 0 | |
| | | | 1, -1, 1 | 0 | |
| | 15 | 12 | 3, 1, 1 | 0 | -0.003609 |
| | | | 1, 1, 1 | 1 | |
| | | | 1, -1, 1 | 0 | |
| | 16 | 24 | 3, 1, 1 | 1 | -0.003316 |
| | | | 1, 1, 1 | 1 | |
| | | | 1, -1, 1 | 0 | |
| | 17 | 12 | 3, 1, 1 | 1 | -0.003237 |
| | | | 1, 1, 1 | 0 | |
| | | | 1, -1, 1 | 1 | |
| | 18 | 12 | 3, 1, 1 | 1 | 0.001410 |
| | | | 1, 1, 1 | 1 | |
| | | | 1, -1, 1 | 1 | |
| | 19 | 8 | 3, 1, 1 | 0 | -0.000988 |
| | | | 1, 1, -1 | 0 | |
| | | | 1, -1, 1 | 0 | |
| | 20 | 24 | 3, 1, 1 | 1 | -0.006162 |
| | | | 1, 1, -1 | 0 | |
| | | | 1, -1, 1 | 0 | |
| | 21 | 24 | 3, 1, 1 | 1 | -0.003798 |
| | | | 1, 1, -1 | 1 | |
| | | | 1, -1, 1 | 0 | |
| | 22 | 8 | 3, 1, 1 | 1 | -0.001938 |
| | | | 1, 1, -1 | 1 | |
| | | | 1, -1, 1 | 1 | |
| | 23 | 24 | 3, 1, 1 | 0 | 0.000989 |
| | | | 2, 2, 2 | 0 | |
| | | | 0, 2, 2 | 0 | |
| | 24 | 24 | 3, 1, 1 | 1 | -0.006418 |
| | | | 2, 2, 2 | 0 | |
| | | | 0, 2, 2 | 0 | |
| | 25 | 24 | 3, 1, 1 | 0 | 0.003440 |
| | | | 2, 2, 2 | 1 | |
| | | | 0, 2, 2 | 0 | |
| | 26 | 24 | 3, 1, 1 | 1 | 0.001680 |
| | | | 2, 2, 2 | 1 | |
| | | | 0, 2, 2 | 0 | |
| | 27 | 24 | 3, 1, 1 | 0 | 0.005852 |
| | | | 2, 2, 2 | 0 | |
| | | | 0, 2, 2 | 1 | |
| | 28 | 24 | 3, 1, 1 | 1 | -0.004249 |
| | | | 2, 2, 2 | 0 | |
| | | | 0, 2, 2 | 1 | |
| | 29 | 24 | 3, 1, 1 | 0 | 0.002548 |
| | | | 2, 2, 2 | 1 | |
| | | | 0, 2, 2 | 1 | |
| | 30 | 24 | 3, 1, 1 | 1 | -0.001349 |
| | | | 2, 2, 2 | 1 | |
| | | | 0, 2, 2 | 1 | |
| | 31 | 48 | 3, 1, 1 | 0 | -0.000536 |
| | | | 2, 0, 2 | 0 | |
| | | | 0, 2, 2 | 0 | |
| | 32 | 48 | 3, 1, 1 | 1 | -0.002730 |
| | | | 2, 0, 2 | 0 | |
| | | | 0, 2, 2 | 0 | |
| | 33 | 48 | 3, 1, 1 | 0 | 0.003826 |
| | | | 2, 0, 2 | 1 | |
| | | | 0, 2, 2 | 0 | |
| | 34 | 48 | 3, 1, 1 | 1 | -0.001826 |
| | | | 2, 0, 2 | 1 | |
| | | | 0, 2, 2 | 0 | |
| | 35 | 48 | 3, 1, 1 | 0 | -0.001100 |
| | | | 2, 0, 2 | 0 | |
| | | | 0, 2, 2 | 1 | |
| | 36 | 48 | 3, 1, 1 | 1 | -0.000005 |
| | | | 2, 0, 2 | 0 | |
| | | | 0, 2, 2 | 1 | |
| | 37 | 48 | 3, 1, 1 | 0 | 0.000020 |
| | | | 2, 0, 2 | 1 | |
| | | | 0, 2, 2 | 1 | |
| | 37 | 48 | 3, 1, 1 | 1 | 0.000732 |
| | | | 2, 0, 2 | 1 | |
| | | | 0, 2, 2 | 1 | |
| | 39 | 24 | 3, 1, 1 | 0 | -0.002413 |
| | | | 3, 1, 3 | 0 | |
| | | | 0, 2, 2 | 0 | |
| | 40 | 48 | 3, 1, 1 | 1 | 0.001855 |
| | | | 3, 1, 3 | 0 | |
| | | | 0, 2, 2 | 0 | |
| | 41 | 24 | 3, 1, 1 | 1 | -0.000673 |
| | | | 3, 1, 3 | 1 | |
| | | | 0, 2, 2 | 0 | |
| | 42 | 24 | 3, 1, 1 | 0 | 0.001396 |
| | | | 3, 1, 3 | 0 | |
| | | | 0, 2, 2 | 1 | |
| | 43 | 48 | 3, 1, 1 | 1 | 0.000772 |
| | | | 3, 1, 3 | 0 | |
| | | | 0, 2, 2 | 1 | |
| | 44 | 24 | 3, 1, 1 | 1 | -0.000940 |
| | | | 3, 1, 3 | 1 | |
| | | | 0, 2, 2 | 1 | |
| | 45 | 24 | 3, 1, 1 | 0 | 0.001739 |
| | | | 1, -1, 1 | 0 | |
| | | | 0, 2, 2 | 0 | |
| | 46 | 48 | 3, 1, 1 | 1 | -0.002150 |
| | | | 1, -1, 1 | 0 | |
| | | | 0, 2, 2 | 0 | |
| | 47 | 24 | 3, 1, 1 | 1 | -0.001545 |
| | | | 1, -1, 1 | 1 | |
| | | | 0, 2, 2 | 0 | |
| | 48 | 24 | 3, 1, 1 | 0 | 0.000802 |
| | | | 1, -1, 1 | 0 | |
| | | | 0, 2, 2 | 1 | |
| | 49 | 48 | 3, 1, 1 | 1 | 0.001607 |
| | | | 1, -1, 1 | 0 | |
| | | | 0, 2, 2 | 1 | |
| | 50 | 24 | 3, 1, 1 | 1 | -0.001042 |
| | | | 1, -1, 1 | 1 | |
| | | | 0, 2, 2 | 1 | |
| | 51 | 12 | 3, 1, 1 | 0 | 0.000187 |
| | | | 3, 3, 3 | 0 | |
| | | | 0, 2, 2 | 0 | |
| | 52 | 24 | 3, 1, 1 | 1 | -0.001974 |
| | | | 3, 3, 3 | 0 | |
| | | | 0, 2, 2 | 0 | |
| | 53 | 12 | 3, 1, 1 | 1 | 0.001771 |
| | | | 3, 3, 3 | 1 | |
| | | | 0, 2, 2 | 0 | |
| | 54 | 12 | 3, 1, 1 | 0 | 0.000105 |
| | | | 3, 3, 3 | 0 | |
| | | | 0, 2, 2 | 1 | |
| | 55 | 24 | 3, 1, 1 | 1 | 0.000230 |
| | | | 3, 3, 3 | 0 | |
| | | | 0, 2, 2 | 1 | |
| | 56 | 12 | 3, 1, 1 | 1 | -0.001423 |
| | | | 3, 3, 3 | 1 | |
| | | | 0, 2, 2 | 1 | |
| | 57 | 24 | 3, 1, 1 | 0 | 0.000411 |
| | | | 4, 2, 2 | 0 | |
| | | | 1, 3, 3 | 0 | |
| | 57 | 24 | 3, 1, 1 | 1 | -0.001910 |
| | | | 4, 2, 2 | 0 | |
| | | | 1, 3, 3 | 0 | |
| | 59 | 24 | 3, 1, 1 | 0 | -0.001062 |
| | | | 4, 2, 2 | 1 | |
| | | | 1, 3, 3 | 0 | |
| | 60 | 24 | 3, 1, 1 | 1 | -0.001043 |
| | | | 4, 2, 2 | 1 | |
| | | | 1, 3, 3 | 0 | |
| | 61 | 24 | 3, 1, 1 | 0 | 0.002388 |
| | | | 4, 2, 2 | 0 | |
| | | | 1, 3, 3 | 1 | |
| | 62 | 24 | 3, 1, 1 | 1 | 0.002410 |
| | | | 4, 2, 2 | 0 | |
| | | | 1, 3, 3 | 1 | |
| | 63 | 24 | 3, 1, 1 | 0 | -0.000769 |
| | | | 4, 2, 2 | 1 | |
| | | | 1, 3, 3 | 1 | |
| | 64 | 24 | 3, 1, 1 | 1 | 0.001858 |
| | | | 4, 2, 2 | 1 | |
| | | | 1, 3, 3 | 1 | |
| | 65 | 4 | 3, 1, 1 | 0 | -0.004558 |
| | | | 2, 2, 2 | 0 | |
| | | | 1, 3, 3 | 0 | |
| | 66 | 8 | 3, 1, 1 | 1 | 0.005351 |
| | | | 4, 2, 2 | 0 | |
| | | | 1, 3, 3 | 0 | |
| | 67 | 4 | 3, 1, 1 | 0 | -0.001518 |
| | | | 4, 2, 2 | 1 | |
| | | | 1, 3, 3 | 0 | |
| | 68 | 8 | 3, 1, 1 | 1 | 0.003486 |
| | | | 4, 2, 2 | 1 | |
| | | | 1, 3, 3 | 0 | |
| | 69 | 4 | 3, 1, 1 | 1 | -0.011909 |
| | | | 4, 2, 2 | 0 | |
| | | | 1, 3, 3 | 1 | |
| | 70 | 4 | 3, 1, 1 | 1 | 0.001619 |
| | | | 4, 2, 2 | 1 | |
| | | | 1, 3, 3 | 1 | |
| | 71 | 24 | 3, 1, 1 | 0 | -0.002856 |
| | | | 1, 1, 1 | 0 | |
| | | | 1, 3, 3 | 0 | |
| | 72 | 24 | 3, 1, 1 | 1 | 0.004765 |
| | | | 1, 1, 1 | 0 | |
| | | | 1, 3, 3 | 0 | |
| | 73 | 48 | 3, 1, 1 | 0 | 0.001230 |
| | | | 1, 1, 1 | 1 | |
| | | | 1, 3, 3 | 0 | |
| | 74 | 24 | 3, 1, 1 | 1 | -0.001436 |
| | | | 1, 1, 1 | 1 | |
| | | | 1, 3, 3 | 0 | |
| | 75 | 24 | 3, 1, 1 | 0 | -0.000758 |
| | | | 1, 1, 1 | 0 | |
| | | | 1, 3, 3 | 1 | |
| | 76 | 24 | 3, 1, 1 | 1 | 0.004072 |
| | | | 1, 1, 1 | 0 | |
| | | | 1, 3, 3 | 1 | |
| | 77 | 24 | 3, 1, 1 | 0 | 0.001395 |
| | | | 1, 1, 1 | 1 | |
| | | | 1, 3, 3 | 1 | |
| | 78 | 24 | 3, 1, 1 | 1 | -0.000319 |
| | | | 1, 1, 1 | 1 | |
| | | | 1, 3, 3 | 1 | |
| | 79 | 24 | 3, 1, 1 | 0 | 0.001988 |
| | | | 0, 0, 2 | 0 | |
| | | | 1, 3, 3 | 0 | |
| | 80 | 48 | 3, 1, 1 | 1 | -0.001710 |
| | | | 0, 0, 2 | 0 | |
| | | | 1, 3, 3 | 0 | |
| | 81 | 24 | 3, 1, 1 | 0 | 0.000355 |
| | | | 0, 0, 2 | 1 | |
| | | | 1, 3, 3 | 0 | |
| | 82 | 48 | 3, 1, 1 | 1 | 0.001244 |
| | | | 0, 0, 2 | 1 | |
| | | | 1, 3, 3 | 0 | |
| | 83 | 24 | 3, 1, 1 | 1 | -0.001068 |
| | | | 0, 0, 2 | 0 | |
| | | | 1, 3, 3 | 1 | |
| | 84 | 24 | 3, 1, 1 | 1 | -0.000762 |
| | | | 0, 0, 2 | 1 | |
| | | | 1, 3, 3 | 1 | |
| quadruplet | 1 | 6 | 3, 1, 1 | 0 | 0.013867 |
| | | | 2, 0, 2 | 0 | |
| | | | 2, 2, 2 | 0 | |
| | | | 1, 1, 1 | 0 | |
| | 2 | 24 | 3, 1, 1 | 1 | -0.008822 |
| | | | 2, 0, 2 | 0 | |
| | | | 2, 2, 2 | 0 | |
| | | | 1, 1, 1 | 0 | |
| | 3 | 24 | 3, 1, 1 | 1 | 0.002717 |
| | | | 2, 0, 2 | 1 | |
| | | | 2, 2, 2 | 0 | |
| | | | 1, 1, 1 | 0 | |
| | 4 | 12 | 3, 1, 1 | 0 | 0.004697 |
| | | | 2, 0, 2 | 1 | |
| | | | 2, 2, 2 | 1 | |
| | | | 1, 1, 1 | 0 | |
| | 5 | 24 | 3, 1, 1 | 1 | -0.001904 |
| | | | 2, 0, 2 | 1 | |
| | | | 2, 2, 2 | 1 | |
| | | | 1, 1, 1 | 0 | |
| | 6 | 6 | 3, 1, 1 | 1 | -0.000504 |
| | | | 2, 0, 2 | 1 | |
| | | | 2, 2, 2 | 1 | |
| | | | 1, 1, 1 | 1 | |
| | 7 | 6 | 3, 1, 1 | 0 | 0.004957 |
| | | | 2, 0, 0 | 0 | |
| | | | 2, 0, 2 | 0 | |
| | | | 1, -1, 1 | 0 | |
| | 8 | 12 | 3, 1, 1 | 1 | 0.008526 |
| | | | 2, 0, 0 | 0 | |
| | | | 2, 0, 2 | 0 | |
| | | | 1, -1, 1 | 0 | |
| | 9 | 12 | 3, 1, 1 | 0 | -0.006806 |
| | | | 2, 0, 0 | 1 | |
| | | | 2, 0, 2 | 0 | |
| | | | 1, -1, 1 | 0 | |
| | 10 | 24 | 3, 1, 1 | 1 | -0.005950 |
| | | | 2, 0, 0 | 1 | |
| | | | 2, 0, 2 | 0 | |
| | | | 1, -1, 1 | 0 | |
| | 11 | 6 | 3, 1, 1 | 0 | -0.000763 |
| | | | 2, 0, 2 | 1 | |
| | | | 2, 2, 2 | 1 | |
| | | | 1, 1, 1 | 0 | |
| | 12 | 12 | 3, 1, 1 | 1 | -0.001082 |
| | | | 2, 0, 2 | 1 | |
| | | | 2, 2, 2 | 1 | |
| | | | 1, 1, 1 | 0 | |
| | 13 | 6 | 3, 1, 1 | 1 | 0.000195 |
| | | | 2, 0, 2 | 0 | |
| | | | 2, 2, 2 | 0 | |
| | | | 1, 1, 1 | 1 | |
| | 14 | 12 | 3, 1, 1 | 1 | 0.002559 |
| | | | 2, 0, 2 | 1 | |
| | | | 2, 2, 2 | 0 | |
| | | | 1, 1, 1 | 1 | |
| | 15 | 6 | 3, 1, 1 | 1 | -0.000442 |
| | | | 2, 0, 2 | 1 | |
| | | | 2, 2, 2 | 1 | |
| | | | 1, 1, 1 | 1 | |
| | 16 | 24 | 3, 1, 1 | 0 | -0.019263 |
| | | | 2, 0, 2 | 0 | |
| | | | 1, 1, 1 | 0 | |
| | | | 1, -1, 1 | 0 | |
| | 17 | 48 | 3, 1, 1 | 1 | 0.001864 |
| | | | 2, 0, 2 | 0 | |
| | | | 1, 1, 1 | 0 | |
| | | | 1, -1, 1 | 0 | |
| | 18 | 24 | 3, 1, 1 | 0 | 0.013411 |
| | | | 2, 0, 2 | 1 | |
| | | | 1, 1, 1 | 0 | |
| | | | 1, -1, 1 | 0 | |
| | 19 | 48 | 3, 1, 1 | 1 | 0.000408 |
| | | | 2, 0, 2 | 1 | |
| | | | 1, 1, 1 | 0 | |
| | | | 1, -1, 1 | 0 | |
| | 20 | 24 | 3, 1, 1 | 0 | 0.008198 |
| | | | 2, 0, 2 | 0 | |
| | | | 1, 1, 1 | 1 | |
| | | | 1, -1, 1 | 0 | |
| | 21 | 48 | 3, 1, 1 | 1 | 0.001353 |
| | | | 2, 0, 2 | 0 | |
| | | | 1, 1, 1 | 1 | |
| | | | 1, -1, 1 | 0 | |
| | 22 | 24 | 3, 1, 1 | 0 | -0.000048 |
| | | | 2, 0, 2 | 1 | |
| | | | 1, 1, 1 | 1 | |
| | | | 1, -1, 1 | 0 | |
| | 23 | 48 | 3, 1, 1 | 1 | -0.003149 |
| | | | 2, 0, 2 | 1 | |
| | | | 1, 1, 1 | 1 | |
| | | | 1, -1, 1 | 0 | |
| | 24 | 24 | 3, 1, 1 | 1 | 0.002273 |
| | | | 2, 0, 2 | 0 | |
| | | | 1, 1, 1 | 0 | |
| | | | 1, -1, 1 | 1 | |
| | 25 | 24 | 3, 1, 1 | 1 | -0.002646 |
| | | | 2, 0, 2 | 1 | |
| | | | 1, 1, 1 | 0 | |
| | | | 1, -1, 1 | 1 | |
| | 26 | 24 | 3, 1, 1 | 1 | 0.007630 |
| | | | 2, 0, 2 | 0 | |
| | | | 1, 1, 1 | 1 | |
| | | | 1, -1, 1 | 1 | |
| | 27 | 24 | 3, 1, 1 | 1 | 0.000307 |
| | | | 2, 0, 2 | 1 | |
| | | | 1, 1, 1 | 1 | |
| | | | 1, -1, 1 | 1 | |
| | 28 | 3 | 3, 1, 1 | 0 | -0.009548 |
| | | | 3, -1, 1 | 0 | |
| | | | 1, 1, 1 | 0 | |
| | | | 1, -1, 1 | 0 | |
| | 29 | 12 | 3, 1, 1 | 1 | 0.000351 |
| | | | 3, -1, 1 | 0 | |
| | | | 1, 1, 1 | 0 | |
| | | | 1, -1, 1 | 0 | |
| | 30 | 12 | 3, 1, 1 | 1 | 0.009039 |
| | | | 3, -1, 1 | 1 | |
| | | | 1, 1, 1 | 0 | |
| | | | 1, -1, 1 | 0 | |
| | 31 | 6 | 3, 1, 1 | 0 | -0.007038 |
| | | | 3, -1, 1 | 1 | |
| | | | 1, 1, 1 | 1 | |
| | | | 1, -1, 1 | 0 | |
| | 32 | 12 | 3, 1, 1 | 1 | -0.004783 |
| | | | 3, -1, 1 | 1 | |
| | | | 1, 1, 1 | 1 | |
| | | | 1, -1, 1 | 0 | |
| | 33 | 3 | 3, 1, 1 | 1 | 0.000449 |
| | | | 3, -1, 1 | 1 | |
| | | | 1, 1, 1 | 1 | |
| | | | 1, -1, 1 | 1 | |
| | 34 | 8 | 3, 1, 1 | 0 | 0.016015 |
| | | | 2, 0, 0 | 0 | |
| | | | 1, 1, -1 | 0 | |
| | | | 1, -1, 1 | 0 | |
| | 35 | 24 | 3, 1, 1 | 1 | -0.0067 |
| | | | 2, 0, 0 | 0 | |
| | | | 1, 1, -1 | 0 | |
| | | | 1, -1, 1 | 0 | |
| | 36 | 8 | 3, 1, 1 | 0 | -0.00946 |
| | | | 2, 0, 0 | 1 | |
| | | | 1, 1, -1 | 0 | |
| | | | 1, -1, 1 | 0 | |
| | 37 | 24 | 3, 1, 1 | 1 | 0.003139 |
| | | | 2, 0, 0 | 1 | |
| | | | 1, 1, -1 | 0 | |
| | | | 1, -1, 1 | 0 | |
| | 38 | 8 | 3, 1, 1 | 1 | -0.00152 |
| | | | 2, 0, 0 | 0 | |
| | | | 1, 1, -1 | 1 | |
| | | | 1, -1, 1 | 0 | |
| | 39 | 24 | 3, 1, 1 | 1 | 0.003412 |
| | | | 2, 0, 0 | 1 | |
| | | | 1, 1, -1 | 1 | |
| | | | 1, -1, 1 | 0 | |
| | 40 | 8 | 3, 1, 1 | 1 | -0.005671 |
| | | | 2, 0, 0 | 0 | |
| | | | 1, 1, -1 | 1 | |
| | | | 1, -1, 1 | 1 | |
| | 41 | 8 | 3, 1, 1 | 1 | -0.000367 |
| | | | 2, 0, 0 | 1 | |
| | | | 1, 1, -1 | 1 | |
| | | | 1, -1, 1 | 1 | |
| | 42 | 8 | 3, 1, 1 | 0 | -0.006003 |
| | | | 1, 1, 1 | 0 | |
| | | | 1, 1, -1 | 0 | |
| | | | 1, -1, 1 | 0 | |
| | 43 | 24 | 3, 1, 1 | 1 | -0.001230 |
| | | | 1, 1, 1 | 0 | |
| | | | 1, 1, -1 | 0 | |
| | | | 1, -1, 1 | 0 | |
| | 44 | 8 | 3, 1, 1 | 0 | 0.010164 |
| | | | 1, 1, 1 | 1 | |
| | | | 1, 1, -1 | 0 | |
| | | | 1, -1, 1 | 0 | |
| | 45 | 24 | 3, 1, 1 | 1 | 0.003687 |
| | | | 1, 1, 1 | 1 | |
| | | | 1, 1, -1 | 0 | |
| | | | 1, -1, 1 | 0 | |
| | 46 | 24 | 3, 1, 1 | 1 | -0.000165 |
| | | | 1, 1, 1 | 0 | |
| | | | 1, 1, -1 | 1 | |
| | | | 1, -1, 1 | 0 | |
| | 47 | 24 | 3, 1, 1 | 1 | 0.000478 |
| | | | 1, 1, 1 | 1 | |
| | | | 1, 1, -1 | 1 | |
| | | | 1, -1, 1 | 0 | |
| | 48 | 8 | 3, 1, 1 | 1 | 0.001911 |
| | | | 1, 1, 1 | 0 | |
| | | | 1, 1, -1 | 1 | |
| | | | 1, -1, 1 | 1 | |
| | 49 | 8 | 3, 1, 1 | 1 | -0.000679 |
| | | | 1, 1, 1 | 1 | |
| | | | 1, 1, -1 | 1 | |
| | | | 1, -1, 1 | 1 | |
| | 50 | 2 | 3, 1, 1 | 0 | -0.013338 |
| | | | 3, -1, -1 | 0 | |
| | | | 1, 1, -1 | 0 | |
| | | | 1, -1, 1 | 0 | |
| | 51 | 8 | 3, 1, 1 | 1 | 0.002936 |
| | | | 3, -1, -1 | 0 | |
| | | | 1, 1, -1 | 0 | |
| | | | 1, -1, 1 | 0 | |
| | 52 | 12 | 3, 1, 1 | 1 | 0.00243 |
| | | | 3, -1, -1 | 1 | |
| | | | 1, 1, -1 | 0 | |
| | | | 1, -1, 1 | 0 | |
| | 53 | 8 | 3, 1, 1 | 1 | 0.002449 |
| | | | 3, -1, -1 | 1 | |
| | | | 1, 1, -1 | 1 | |
| | | | 1, -1, 1 | 0 | |
| | 54 | 2 | 3, 1, 1 | 1 | -0.00032 |
| | | | 3, -1, -1 | 1 | |
| | | | 1, 1, -1 | 1 | |
| | | | 1, -1, 1 | 1 | |
| | 55 | 48 | 3, 1, 1 | 0 | 0.00951 |
| | | | 2, 0, 2 | 0 | |
| | | | 2, 2, 2 | 0 | |
| | | | 0, 2, 2 | 0 | |
| | 56 | 48 | 3, 1, 1 | 1 | -0.004389 |
| | | | 2, 0, 2 | 0 | |
| | | | 2, 2, 2 | 0 | |
| | | | 0, 2, 2 | 0 | |
| | 57 | 48 | 3, 1, 1 | 0 | -0.005972 |
| | | | 2, 0, 2 | 1 | |
| | | | 2, 2, 2 | 0 | |
| | | | 0, 2, 2 | 0 | |
| | 58 | 48 | 3, 1, 1 | 1 | -0.000022 |
| | | | 2, 0, 2 | 1 | |
| | | | 2, 2, 2 | 0 | |
| | | | 0, 2, 2 | 0 | |
| | 59 | 48 | 3, 1, 1 | 0 | -0.004206 |
| | | | 2, 0, 2 | 0 | |
| | | | 2, 2, 2 | 1 | |
| | | | 0, 2, 2 | 0 | |
| | 60 | 48 | 3, 1, 1 | 1 | -0.000617 |
| | | | 2, 0, 2 | 0 | |
| | | | 2, 2, 2 | 1 | |
| | | | 0, 2, 2 | 0 | |
| | 61 | 48 | 3, 1, 1 | 0 | -0.000196 |
| | | | 2, 0, 2 | 1 | |
| | | | 2, 2, 2 | 1 | |
| | | | 0, 2, 2 | 0 | |
| | 62 | 48 | 3, 1, 1 | 1 | 0.003091 |
| | | | 2, 0, 2 | 1 | |
| | | | 2, 2, 2 | 1 | |
| | | | 0, 2, 2 | 0 | |
| | 63 | 48 | 3, 1, 1 | 0 | 0.002709 |
| | | | 2, 0, 2 | 0 | |
| | | | 2, 2, 2 | 0 | |
| | | | 0, 2, 2 | 1 | |
| | 64 | 48 | 3, 1, 1 | 1 | -0.001293 |
| | | | 2, 0, 2 | 0 | |
| | | | 2, 2, 2 | 0 | |
| | | | 0, 2, 2 | 1 | |
| | 65 | 48 | 3, 1, 1 | 0 | -0.000739 |
| | | | 2, 0, 2 | 1 | |
| | | | 2, 2, 2 | 0 | |
| | | | 0, 2, 2 | 1 | |
| | 66 | 48 | 3, 1, 1 | 1 | 0.000747 |
| | | | 2, 0, 2 | 1 | |
| | | | 2, 2, 2 | 0 | |
| | | | 0, 2, 2 | 1 | |
| | 67 | 48 | 3, 1, 1 | 0 | -0.002849 |
| | | | 2, 0, 2 | 0 | |
| | | | 2, 2, 2 | 1 | |
| | | | 0, 2, 2 | 1 | |
| | 68 | 48 | 3, 1, 1 | 1 | 0.002368 |
| | | | 2, 0, 2 | 0 | |
| | | | 2, 2, 2 | 1 | |
| | | | 0, 2, 2 | 1 | |
| | 69 | 48 | 3, 1, 1 | 0 | -0.003896 |
| | | | 2, 0, 2 | 1 | |
| | | | 2, 2, 2 | 1 | |
| | | | 0, 2, 2 | 1 | |
| | 70 | 48 | 3, 1, 1 | 1 | -0.000356 |
| | | | 2, 0, 2 | 1 | |
| | | | 2, 2, 2 | 1 | |
| | | | 0, 2, 2 | 1 | |
| | 71 | 24 | 3, 1, 1 | 0 | -0.006916 |
| | | | 2, 2, 0 | 0 | |
| | | | 2, 0, 2 | 0 | |
| | | | 0, 2, 2 | 0 | |
| | 72 | 24 | 3, 1, 1 | 1 | 0.001142 |
| | | | 2, 2, 0 | 0 | |
| | | | 2, 0, 2 | 0 | |
| | | | 0, 2, 2 | 0 | |
| | 73 | 48 | 3, 1, 1 | 0 | 0.008173 |
| | | | 2, 2, 0 | 1 | |
| | | | 2, 0, 2 | 0 | |
| | | | 0, 2, 2 | 0 | |
| | 74 | 24 | 3, 1, 1 | 1 | -0.001360 |
| | | | 2, 2, 0 | 1 | |
| | | | 2, 0, 2 | 0 | |
| | | | 0, 2, 2 | 0 | |
| | 75 | 24 | 3, 1, 1 | 0 | -0.001421 |
| | | | 2, 2, 0 | 1 | |
| | | | 2, 0, 2 | 1 | |
| | | | 0, 2, 2 | 0 | |
| | 76 | 24 | 3, 1, 1 | 1 | -0.001643 |
| | | | 2, 2, 0 | 1 | |
| | | | 2, 0, 2 | 1 | |
| | | | 0, 2, 2 | 0 | |
| | 77 | 24 | 3, 1, 1 | 0 | -0.009752 |
| | | | 2, 2, 0 | 0 | |
| | | | 2, 0, 2 | 0 | |
| | | | 0, 2, 2 | 1 | |
| | 78 | 24 | 3, 1, 1 | 1 | 0.002740 |
| | | | 2, 2, 0 | 0 | |
| | | | 2, 0, 2 | 0 | |
| | | | 0, 2, 2 | 1 | |
| | 79 | 48 | 3, 1, 1 | 0 | 0.003431 |
| | | | 2, 2, 0 | 1 | |
| | | | 2, 0, 2 | 0 | |
| | | | 0, 2, 2 | 1 | |
| | 80 | 48 | 3, 1, 1 | 1 | -0.001030 |
| | | | 2, 2, 0 | 1 | |
| | | | 2, 0, 2 | 0 | |
| | | | 0, 2, 2 | 1 | |
| | 81 | 24 | 3, 1, 1 | 0 | 0.001332 |
| | | | 2, 2, 0 | 1 | |
| | | | 2, 0, 2 | 1 | |
| | | | 0, 2, 2 | 1 | |
| | 82 | 24 | 3, 1, 1 | 1 | 0.000696 |
| | | | 2, 2, 0 | 1 | |
| | | | 2, 0, 2 | 1 | |
| | | | 0, 2, 2 | 1 | |
| | 83 | 12 | 3, 1, 1 | 0 | -0.008776 |
| | | | 2, 2, 2 | 0 | |
| | | | 1, 1, 1 | 0 | |
| | | | 0, 2, 2 | 0 | |
| | 84 | 24 | 3, 1, 1 | 1 | -0.001942 |
| | | | 2, 2, 2 | 0 | |
| | | | 1, 1, 1 | 0 | |
| | | | 0, 2, 2 | 0 | |
| | 85 | 24 | 3, 1, 1 | 0 | 0.007327 |
| | | | 2, 2, 2 | 1 | |
| | | | 1, 1, 1 | 0 | |
| | | | 0, 2, 2 | 0 | |
| | 86 | 12 | 3, 1, 1 | 1 | 0.001902 |
| | | | 2, 2, 2 | 1 | |
| | | | 1, 1, 1 | 0 | |
| | | | 0, 2, 2 | 0 | |
| | 87 | 12 | 3, 1, 1 | 1 | -0.002243 |
| | | | 2, 2, 2 | 0 | |
| | | | 1, 1, 1 | 1 | |
| | | | 0, 2, 2 | 0 | |
| | 88 | 12 | 3, 1, 1 | 0 | -0.003588 |
| | | | 2, 2, 2 | 1 | |
| | | | 1, 1, 1 | 1 | |
| | | | 0, 2, 2 | 0 | |
| | 89 | 24 | 3, 1, 1 | 1 | 0.002440 |
| | | | 2, 2, 2 | 1 | |
| | | | 1, 1, 1 | 1 | |
| | | | 0, 2, 2 | 0 | |
| | 90 | 12 | 3, 1, 1 | 1 | -0.003050 |
| | | | 2, 2, 2 | 0 | |
| | | | 1, 1, 1 | 0 | |
| | | | 0, 2, 2 | 1 | |
| | 91 | 24 | 3, 1, 1 | 1 | 0.000048 |
| | | | 2, 2, 2 | 1 | |
| | | | 1, 1, 1 | 0 | |
| | | | 0, 2, 2 | 1 | |
| | 92 | 12 | 3, 1, 1 | 1 | -0.000170 |
| | | | 2, 2, 2 | 1 | |
| | | | 1, 1, 1 | 1 | |
| | | | 0, 2, 2 | 1 | |
| | 93 | 48 | 3, 1, 1 | 0 | -0.000927 |
| | | | 2, 0, 2 | 0 | |
| | | | 1, 1, 1 | 0 | |
| | | | 0, 2, 2 | 0 | |
| | 94 | 48 | 3, 1, 1 | 1 | 0.003440 |
| | | | 2, 0, 2 | 0 | |
| | | | 1, 1, 1 | 0 | |
| | | | 0, 2, 2 | 0 | |
| | 95 | 48 | 3, 1, 1 | 0 | -0.000250 |
| | | | 2, 0, 2 | 1 | |
| | | | 1, 1, 1 | 0 | |
| | | | 0, 2, 2 | 0 | |
| | 96 | 48 | 3, 1, 1 | 1 | -0.001796 |
| | | | 2, 0, 2 | 1 | |
| | | | 1, 1, 1 | 0 | |
| | | | 0, 2, 2 | 0 | |
| | 97 | 48 | 3, 1, 1 | 0 | -0.001864 |
| | | | 2, 0, 2 | 0 | |
| | | | 1, 1, 1 | 1 | |
| | | | 0, 2, 2 | 0 | |
| | 98 | 48 | 3, 1, 1 | 1 | -0.000651 |
| | | | 2, 0, 2 | 0 | |
| | | | 1, 1, 1 | 1 | |
| | | | 0, 2, 2 | 0 | |
| | 99 | 48 | 3, 1, 1 | 0 | 0.001080 |
| | | | 2, 0, 2 | 1 | |
| | | | 1, 1, 1 | 1 | |
| | | | 0, 2, 2 | 0 | |
| | 100 | 48 | 3, 1, 1 | 1 | -0.001456 |
| | | | 2, 0, 2 | 1 | |
| | | | 1, 1, 1 | 1 | |
| | | | 0, 2, 2 | 0 | |

**Table S3.** Cluster expansion for $CaF_2$ lattice.

| cluster | labels | multiplicity | coordinates | point function index | ECI |
|---|---|---|---|---|---|
| empty | 1 | 1 | | | -0.860389 |
| point | 1 | 2 | 0.25, 0.25, 0.25 | 0 | -0.071377 |
| | 2 | 2 | 0.25, 0.25, 0.25 | 1 | 0.119210 |
| | 3 | 1 | 1, 1, 1 | 0 | 0.036823 |
| | 4 | 1 | 1, 1, 1 | 1 | -0.009440 |
| pair | 1 | 2 | 1, 1, 1 | 0 | 0.045450 |
| | | | 1.25, 1.25, 1.25 | 0 | |
| | 2 | 2 | 1, 1, 1 | 1 | -0.295540 |
| | | | 1.25, 1.25, 1.25 | 0 | |
| | 3 | 2 | 1, 1, 1 | 0 | -0.054460 |
| | | | 1.25, 1.25, 1.25 | 1 | |
| | 4 | 2 | 1, 1, 1 | 1 | 0.033520 |
| | | | 1.25, 1.25, 1.25 | 1 | |
| | 5 | 6 | 1, 1, 1 | 0 | 0.047330 |
| | | | 1.25, 1.25, 0.25 | 0 | |
| | 6 | 6 | 1, 1, 1 | 1 | 0.044080 |
| | | | 1.25, 1.25, 0.25 | 0 | |
| | 7 | 6 | 1, 1, 1 | 0 | -0.027299 |
| | | | 1.25, 1.25, 0.25 | 1 | |
| | 8 | 6 | 1, 1, 1 | 1 | 0.008665 |
| | | | 1.25, 1.25, 0.25 | 1 | |
| | 9 | 3 | 0.25, 0.25, 0.25 | 0 | 0.082230 |
| | | | 0.75, 0.75, -0.25 | 0 | |
| | 10 | 6 | 0.25, 0.25, 0.25 | 1 | -0.035747 |
| | | | 0.75, 0.75, -0.25 | 0 | |
| | 11 | 3 | 0.25, 0.25, 0.25 | 1 | -0.020457 |
| | | | 0.75, 0.75, -0.25 | 1 | |
| | 12 | 3 | 0.25, 0.25, 0.25 | 0 | 0.050243 |
| | | | -0.25, -0.25, 0.75 | 0 | |
| | 13 | 6 | 0.25, 0.25, 0.25 | 1 | -0.020095 |
| | | | -0.25, -0.25, 0.75 | 0 | |
| | 14 | 3 | 0.25, 0.25, 0.25 | 1 | 0.072595 |
| | | | -0.25, -0.25, 0.75 | 1 | |
| | 15 | 3 | 1, 1, 1 | 0 | 0.065526 |
| | | | 1, 1, 0 | 0 | |
| | 16 | 6 | 1, 1, 1 | 1 | 0.030908 |
| | | | 1, 1, 0 | 0 | |
| | 17 | 3 | 1, 1, 1 | 1 | -0.026738 |
| | | | 1, 1, 0 | 1 | |
| | 18 | 6 | 0.75, 0.75, 0.75 | 0 | -0.009802 |
| | | | 0.75, 0.75, -0.25 | 0 | |
| | 19 | 6 | 0.75, 0.75, 0.75 | 1 | 0.084155 |
| | | | 0.75, 0.75, -0.25 | 0 | |
| | 20 | 6 | 0.75, 0.75, 0.75 | 1 | -0.034920 |
| | | | 0.75, 0.75, -0.25 | 1 | |
| | 21 | 3 | 1, 1, 1 | 0 | 0.001148 |
| | | | 2, 1, 0 | 0 | |
| | 22 | 6 | 1, 1, 1 | 1 | -0.019640 |
| | | | 2, 1, 0 | 0 | |
| | 23 | 3 | 1, 1, 1 | 1 | 0.005370 |
| | | | 2, 1, 0 | 1 | |
| | 24 | 6 | 0.75, 0.75, 0.75 | 0 | 0.008577 |
| | | | 1.75, 0.75, -0.25 | 0 | |
| | 25 | 12 | 0.75, 0.75, 0.75 | 1 | -0.011669 |
| | | | 1.75, 0.75, -0.25 | 0 | |
| | 26 | 6 | 0.75, 0.75, 0.75 | 1 | 0.040283 |
| | | | 1.75, 0.75, -0.25 | 1 | |
| | 27 | 6 | 0.25, 0.25, 0.25 | 0 | 0.003848 |
| | | | 1, 0, 1 | 0 | |
| | 28 | 6 | 0.25, 0.25, 0.25 | 1 | -0.017097 |
| | | | 1, 0, 1 | 0 | |
| | 29 | 6 | 0.25, 0.25, 0.25 | 0 | 0.009584 |
| | | | 1, 0, 1 | 1 | |
| | 30 | 6 | 0.25, 0.25, 0.25 | 1 | -0.010064 |
| | | | 1, 0, 1 | 1 | |
| | 31 | 6 | 1, 1, 1 | 0 | 0.016909 |
| | | | 1.75, 1.75, -0.25 | 0 | |
| | 32 | 6 | 1, 1, 1 | 1 | -0.069844 |
| | | | 1.75, 1.75, -0.25 | 0 | |
| | 33 | 6 | 1, 1, 1 | 0 | 0.009263 |
| | | | 1.75, 1.75, -0.25 | 1 | |
| | 34 | 6 | 1, 1, 1 | 1 | 0.025537 |
| | | | 1.75, 1.75, -0.25 | 1 | |
| | 35 | 1 | 0.25, 0.25, 0.25 | 0 | -16.656353 |
| | | | -0.25, -0.25, -0.25 | 0 | |
| | 36 | 2 | 0.25, 0.25, 0.25 | 1 | -0.067377 |
| | | | -0.25, -0.25, -0.25 | 0 | |
| | 37 | 1 | 0.25, 0.25, 0.25 | 1 | 0.036861 |
| | | | -0.25, -0.25, -0.25 | 1 | |
| | 38 | 1 | 0.75, 0.75, 0.75 | 0 | -0.144668 |
| | | | 0.25, 0.25, 0.25 | 0 | |
| | 39 | 2 | 0.75, 0.75, 0.75 | 1 | 0.039272 |
| | | | 0.25, 0.25, 0.25 | 0 | |
| | 40 | 1 | 0.75, 0.75, 0.75 | 1 | 0.008002 |
| | | | 0.25, 0.25, 0.25 | 1 | |
| | 41 | 3 | 0.75, 0.75, 0.75 | 0 | 5.596519 |
| | | | 1.25, 1.25, -0.75 | 0 | |
| | 42 | 3 | 1, 1, 1 | 0 | 0.002048 |
| | | | 2, 2, 0 | 0 | |
| | 43 | 6 | 1, 1, 1 | 1 | 0.023137 |
| | | | 2, 2, 0 | 0 | |
| | 44 | 3 | 1, 1, 1 | 1 | 0.027824 |
| | | | 2, 2, 0 | 1 | |
| | 45 | 6 | 0.75, 0.75, 0.75 | 0 | 0.043585 |
| | | | 1.75, 1.75, -0.25 | 0 | |
| | 46 | 6 | 0.75, 0.75, 0.75 | 1 | -0.031648 |
| | | | 1.75, 1.75, -0.25 | 0 | |
| | 47 | 6 | 0.75, 0.75, 0.75 | 1 | 0.052210 |
| | | | 1.75, 1.75, -0.25 | 1 | |
| triplet | 1 | 6 | 0.25, 0.25, 0.25 | 0 | -0.006398 |
| | | | 1, 0, 0 | 0 | |
| | | | 0.75, 0.75, -0.25 | 0 | |
| | 2 | 12 | 0.25, 0.25, 0.25 | 1 | -0.011735 |
| | | | 1, 0, 0 | 0 | |
| | | | 0.75, 0.75, -0.25 | 0 | |
| | 3 | 6 | 0.25, 0.25, 0.25 | 0 | 0.054607 |
| | | | 1, 0, 0 | 1 | |
| | | | 0.75, 0.75, -0.25 | 0 | |
| | 4 | 12 | 0.25, 0.25, 0.25 | 1 | 0.000661 |
| | | | 1, 0, 0 | 1 | |
| | | | 0.75, 0.75, -0.25 | 0 | |
| | 5 | 6 | 0.25, 0.25, 0.25 | 1 | -0.007893 |
| | | | 1, 0, 0 | 0 | |
| | | | 0.75, 0.75, -0.25 | 1 | |
| | 6 | 6 | 0.25, 0.25, 0.25 | 1 | -0.049247 |
| | | | 1, 0, 0 | 1 | |
| | | | 0.75, 0.75, -0.25 | 1 | |
| | 7 | 6 | 0.25, 0.25, 0.25 | 0 | -0.007379 |
| | | | 0, 0, 0 | 0 | |
| | | | -0.25, -0.25, 0.75 | 0 | |
| | 8 | 6 | 0.25, 0.25, 0.25 | 1 | -0.016140 |
| | | | 0, 0, 0 | 0 | |
| | | | -0.25, -0.25, 0.75 | 0 | |
| | 9 | 6 | 0.25, 0.25, 0.25 | 0 | -0.016816 |
| | | | 0, 0, 0 | 1 | |
| | | | -0.25, -0.25, 0.75 | 0 | |
| | 10 | 6 | 0.25, 0.25, 0.25 | 1 | -0.066521 |
| | | | 0, 0, 0 | 1 | |
| | | | -0.25, -0.25, 0.75 | 0 | |
| | 11 | 6 | 0.25, 0.25, 0.25 | 0 | 0.013262 |
| | | | 0, 0, 0 | 0 | |
| | | | -0.25, -0.25, 0.75 | 1 | |
| | 12 | 6 | 0.25, 0.25, 0.25 | 1 | 0.003099 |
| | | | 0, 0, 0 | 0 | |
| | | | -0.25, -0.25, 0.75 | 1 | |
| | 13 | 6 | 0.25, 0.25, 0.25 | 0 | 0.069826 |
| | | | 0, 0, 0 | 1 | |
| | | | -0.25, -0.25, 0.75 | 1 | |
| | 14 | 6 | 0.25, 0.25, 0.25 | 1 | 0.054617 |
| | | | 0, 0, 0 | 1 | |
| | | | -0.25, -0.25, 0.75 | 1 | |
| | 15 | 6 | 1, 1, 1 | 0 | 0.036071 |
| | | | 0.75, 0.75, 0.75 | 0 | |
| | | | 1, 1, 0 | 0 | |
| | 16 | 6 | 1, 1, 1 | 1 | 0.047419 |
| | | | 0.75, 0.75, 0.75 | 0 | |
| | | | 1, 1, 0 | 0 | |
| | 17 | 6 | 1, 1, 1 | 0 | -0.017662 |
| | | | 0.75, 0.75, 0.75 | 1 | |
| | | | 1, 1, 0 | 0 | |
| | 18 | 6 | 1, 1, 1 | 1 | -0.005994 |
| | | | 0.75, 0.75, 0.75 | 1 | |
| | | | 1, 1, 0 | 0 | |
| | 19 | 6 | 1, 1, 1 | 0 | 0.013226 |
| | | | 0.75, 0.75, 0.75 | 0 | |
| | | | 1, 1, 0 | 1 | |
| | 20 | 6 | 1, 1, 1 | 1 | 0.007731 |
| | | | 0.75, 0.75, 0.75 | 0 | |
| | | | 1, 1, 0 | 1 | |
| | 21 | 6 | 1, 1, 1 | 0 | -0.067213 |
| | | | 0.75, 0.75, 0.75 | 1 | |
| | | | 1, 1, 0 | 1 | |
| | 22 | 6 | 1, 1, 1 | 1 | 0.026724 |
| | | | 0.75, 0.75, 0.75 | 1 | |
| | | | 1, 1, 0 | 1 | |
| | 23 | 6 | 0.75, 0.75, 0.75 | 0 | -0.000835 |
| | | | 1, 1, 0 | 0 | |
| | | | 0.75, 0.75, -0.25 | 0 | |
| | 24 | 6 | 0.75, 0.75, 0.75 | 1 | 0.034636 |
| | | | 1, 1, 0 | 0 | |
| | | | 0.75, 0.75, -0.25 | 0 | |
| | 25 | 6 | 0.75, 0.75, 0.75 | 0 | -0.012782 |
| | | | 1, 1, 0 | 1 | |
| | | | 0.75, 0.75, -0.25 | 0 | |
| | 26 | 6 | 0.75, 0.75, 0.75 | 1 | 0.024287 |
| | | | 1, 1, 0 | 1 | |
| | | | 0.75, 0.75, -0.25 | 0 | |
| | 27 | 6 | 0.75, 0.75, 0.75 | 0 | -0.023473 |
| | | | 1, 1, 0 | 0 | |
| | | | 0.75, 0.75, -0.25 | 1 | |
| | 28 | 6 | 0.75, 0.75, 0.75 | 1 | 0.053544 |
| | | | 1, 1, 0 | 0 | |
| | | | 0.75, 0.75, -0.25 | 1 | |
| | 29 | 6 | 0.75, 0.75, 0.75 | 0 | 0.000637 |
| | | | 1, 1, 0 | 1 | |
| | | | 0.75, 0.75, -0.25 | 1 | |
| | 30 | 6 | 0.75, 0.75, 0.75 | 1 | 0.046850 |
| | | | 1, 1, 0 | 1 | |
| | | | 0.75, 0.75, -0.25 | 1 | |
| | 31 | 12 | 0.75, 0.75, 0.75 | 0 | 0.035175 |
| | | | 1.25,0.25, 0.25 | 0 | |
| | | | 0.75, 0.75, -0.25 | 0 | |
| | 32 | 12 | 0.75, 0.75, 0.75 | 1 | 0.006026 |
| | | | 1.25,0.25, 0.25 | 0 | |
| | | | 0.75, 0.75, -0.25 | 0 | |
| | 33 | 12 | 0.75, 0.75, 0.75 | 0 | -0.026214 |
| | | | 1.25,0.25, 0.25 | 1 | |
| | | | 0.75, 0.75, -0.25 | 0 | |
| | 34 | 12 | 0.75, 0.75, 0.75 | 1 | -0.026561 |
| | | | 1.25,0.25, 0.25 | 1 | |
| | | | 0.75, 0.75, -0.25 | 0 | |

**Table S4.** Cluster expansion for bct lattice.

| cluster | labels | multiplicity | coordinates | point function index | ECI |
|---|---|---|---|---|---|
| empty | 1 | 1 | | | -1.710449 |
| point | 1 | 4 | 0.658916, 0.182169, 0.5 | 0 | 0.135773 |
| | 2 | 4 | 0.658916, 0.182169, 0.5 | 1 | 0.190819 |
| | 3 | 2 | 0.25, 1, 1 | 0 | 0.434138 |
| | 4 | 2 | 0.25, 1, 1 | 1 | -0.363188 |
| pair | 1 | 2 | 0.75, 1, 1 | 0 | 0.026428 |
| | | | 1.25, 1, 1 | 0 | |
| | 2 | 4 | 0.75, 1, 1 | 1 | -0.044448 |
| | | | 1.25, 1, 1 | 0 | |
| | 3 | 2 | 0.75, 1, 1 | 1 | 0.071688 |
| | | | 1.25, 1, 1 | 1 | |
| | 4 | 16 | 0.25, 1, 1 | 0 | 0.063500 |
| | | | 0.341084, 0.817831, 0.5 | 0 | |
| | 5 | 16 | 0.25, 1, 1 | 1 | 0.153677 |
| | | | 0.341084, 0.817831, 0.5 | 0 | |
| | 6 | 16 | 0.25, 1, 1 | 0 | 0.018252 |
| | | | 0.341084, 0.817831, 0.5 | 1 | |
| | 7 | 16 | 0.25, 1, 1 | 1 | 0.242494 |
| | | | 0.341084, 0.817831, 0.5 | 1 | |
| | 8 | 2 | 0.158916, 0.5, 0.182169 | 0 | 0.323810 |
| | | | -0.158915, 0.5, 0.817831 | 0 | |
| | 9 | 4 | 0.158916, 0.5, 0.182169 | 1 | 0.022565 |
| | | | -0.158915, 0.5, 0.817831 | 0 | |
| | 10 | 2 | 0.158916, 0.5, 0.182169 | 1 | -0.496630 |
| | | | -0.158915, 0.5, 0.817831 | 1 | |
| | 11 | 4 | 0.158916, 0.5, 0.182169 | 0 | 0.039560 |
| | | | -0.158915, 0.5, -0.182169 | 0 | |
| | 12 | 8 | 0.158916, 0.5, 0.182169 | 1 | -0.029266 |
| | | | -0.158915, 0.5, -0.182169 | 0 | |
| | 13 | 4 | 0.158916, 0.5, 0.182169 | 1 | 0.079151 |
| | | | -0.158915, 0.5, -0.182169 | 1 | |
| | 14 | 8 | 0.341084, 0.817831, 0.5 | 0 | 0.032828 |
| | | | -0.158915, 0.5, 0.817831 | 0 | |
| | 15 | 16 | 0.341084, 0.817831, 0.5 | 1 | -0.018505 |
| | | | -0.158915, 0.5, 0.817831 | 0 | |
| | 16 | 8 | 0.341084, 0.817831, 0.5 | 1 | 0.006927 |
| | | | -0.158915, 0.5, 0.817831 | 1 | |
| | 17 | 8 | 0.158916, 0.5, 0.182169 | 0 | -0.014855 |
| | | | 0.341084, 0.817831, -0.5 | 0 | |
| | 18 | 8 | 0.158916, 0.5, 0.182169 | 1 | 0.032655 |
| | | | 0.341084, 0.817831, -0.5 | 0 | |
| | 19 | 8 | 0.158916, 0.5, 0.182169 | 0 | 0.068584 |
| | | | 0.341084, 0.817831, -0.5 | 1 | |
| | 20 | 8 | 0.158916, 0.5, 0.182169 | 1 | -0.040845 |
| | | | 0.341084, 0.817831, -0.5 | 1 | |
| | 21 | 16 | 0.341084, 0.817831, 0.5 | 0 | 0.058537 |
| | | | 0.75, 0, 1 | 0 | |
| | 22 | 16 | 0.341084, 0.817831, 0.5 | 1 | 0.217077 |
| | | | 0.75, 0, 1 | 0 | |
| | 23 | 16 | 0.341084, 0.817831, 0.5 | 0 | -0.026349 |
| | | | 0.75, 0, 1 | 1 | |
| | 24 | 16 | 0.341084, 0.817831, 0.5 | 1 | -0.092562 |
| | | | 0.75, 0, 1 | 1 | |
| | 25 | 4 | 0.75, 1, 1 | 0 | 0.054751 |
| | | | 1.25, 0, 1 | 0 | |
| | 26 | 8 | 0.75, 1, 1 | 1 | 0.033020 |
| | | | 1.25, 0, 1 | 0 | |
| | 27 | 4 | 0.75, 1, 1 | 1 | -0.001802 |
| | | | 1.25, 0, 1 | 1 | |
| | 28 | 16 | 0.75, 1, 1 | 0 | -0.014362 |
| | | | 0.341084, 0.817831, 0.5 | 0 | |
| | 29 | 16 | 0.75, 1, 1 | 1 | -0.173067 |
| | | | 0.341084, 0.817831, 0.5 | 0 | |
| | 30 | 16 | 0.75, 1, 1 | 0 | -0.024840 |
| | | | 0.341084, 0.817831, 0.5 | 1 | |
| | 31 | 16 | 0.75, 1, 1 | 1 | -0.220632 |
| | | | 0.341084, 0.817831, 0.5 | 1 | |
| | 32 | 4 | 0.658916, 0.182169, 0.5 | 0 | 0.021599 |
| | | | 1.341084, -0.182169, -0.5 | 0 | |
| | 33 | 8 | 0.658916, 0.182169, 0.5 | 1 | -0.029639 |
| | | | 1.341084, -0.182169, -0.5 | 0 | |
| | 34 | 4 | 0.658916, 0.182169, 0.5 | 1 | 0.001830 |
| | | | 1.341084, -0.182169, -0.5 | 1 | |
| | 35 | 8 | 0.158916, 0.5, 0.182169 | 0 | 0.014802 |
| | | | 0.341084, -0.182169, -0.5 | 0 | |
| | 36 | 16 | 0.158916, 0.5, 0.182169 | 1 | -0.003932 |
| | | | 0.341084, -0.182169, -0.5 | 0 | |
| | 37 | 8 | 0.158916, 0.5, 0.182169 | 1 | 0.052789 |
| | | | 0.341084, -0.182169, -0.5 | 1 | |
| | 38 | 2 | 0.75, 1, 1 | 0 | 0.350805 |
| | | | -0.25, 1, 1 | 0 | |
| | 39 | 4 | 0.75, 1, 1 | 1 | 0.624464 |
| | | | -0.25, 1, 1 | 0 | |
| | 40 | 2 | 0.75, 1, 1 | 1 | -0.486807 |
| | | | -0.25, 1, 1 | 1 | |
| | 41 | 4 | 0.341084, 0.817831, 0.5 | 0 | 0.064744 |
| | | | -0.658916, 0.817831, 0.5 | 0 | |
| | 42 | 8 | 0.341084, 0.817831, 0.5 | 1 | -0.050474 |
| | | | -0.658916, 0.817831, 0.5 | 0 | |
| | 43 | 4 | 0.341084, 0.817831, 0.5 | 1 | -0.051766 |
| | | | -0.658916, 0.817831, 0.5 | 1 | |
| | 44 | 8 | 0.841085, 0.5, 0.817831 | 0 | 0.019237 |
| | | | 0.841085, 0.5, -0.182169 | 0 | |
| | 45 | 8 | 0.841085, 0.5, 0.817831 | 1 | 0.011719 |
| | | | 0.841085, 0.5, -0.182169 | 0 | |
| | 46 | 8 | 0.841085, 0.5, 0.817831 | 1 | -0.010851 |
| | | | 0.841085, 0.5, -0.182169 | 1 | |
| | 47 | 8 | 0.341084, 0.817831, 0.5 | 0 | 0.050154 |
| | | | 0.341084, 0.817831, 1.50 | 0 | |
| | 48 | 16 | 0.341084, 0.817831, 0.5 | 1 | -0.033953 |
| | | | 0.341084, 0.817831, 1.50 | 0 | |
| | 49 | 8 | 0.341084, 0.817831, 0.5 | 1 | 0.017057 |
| | | | 0.341084, 0.817831, 1.50 | 1 | |
| | 50 | 8 | 0.75, 1, 1 | 0 | 0.027346 |
| | | | 0.75, 0, 1 | 0 | |
| | 51 | 16 | 0.75, 1, 1 | 1 | 0.001659 |
| | | | 0.75, 0, 1 | 0 | |
| | 52 | 8 | 0.75, 1, 1 | 1 | 0.028487 |
| | | | 0.75, 0, 1 | 1 | |
| | 53 | 16 | 0.25, 1, 1 | 0 | -0.041358 |
| | | | -0.658916, 1.817831, 0.5 | 0 | |
| | 54 | 16 | 0.25, 1, 1 | 1 | 0.013674 |
| | | | -0.658916, 1.817831, 0.5 | 0 | |
| | 55 | 16 | 0.25, 1, 1 | 0 | -0.225012 |
| | | | -0.658916, 1.817831, 0.5 | 1 | |
| | 56 | 16 | 0.25, 1, 1 | 1 | 0.082232 |
| | | | -0.658916, 1.817831, 0.5 | 1 | |
| | 57 | 4 | 0.841085, 0.5, 0.817831 | 0 | -0.030692 |
| | | | 2.158916, 0.5, 0.182169 | 0 | |
| | 58 | 8 | 0.841085, 0.5, 0.817831 | 1 | -0.002911 |
| | | | 2.158916, 0.5, 0.182169 | 0 | |
| | 59 | 4 | 0.841085, 0.5, 0.817831 | 1 | 0.271010 |
| | | | 2.158916, 0.5, 0.182169 | 1 | |
| triplet | 1 | 8 | 0.25, 1, 1 | 0 | 0.032574 |
| | | | -0.25, 1, 1 | 0 | |
| | | | 0.341084, 0.817831, 0.5 | 0 | |
| | 2 | 16 | 0.25, 1, 1 | 1 | -0.001738 |
| | | | -0.25, 1, 1 | 0 | |
| | | | 0.341084, 0.817831, 0.5 | 0 | |
| | 3 | 8 | 0.25, 1, 1 | 1 | 0.004526 |
| | | | -0.25, 1, 1 | 1 | |
| | | | 0.341084, 0.817831, 0.5 | 0 | |
| | 4 | 8 | 0.25, 1, 1 | 0 | -0.033708 |
| | | | -0.25, 1, 1 | 0 | |
| | | | 0.341084, 0.817831, 0.5 | 1 | |
| | 5 | 16 | 0.25, 1, 1 | 1 | -0.000693 |
| | | | -0.25, 1, 1 | 0 | |
| | | | 0.341084, 0.817831, 0.5 | 1 | |
| | 6 | 8 | 0.25, 1, 1 | 1 | -0.024128 |
| | | | -0.25, 1, 1 | 1 | |
| | | | 0.341084, 0.817831, 0.5 | 1 | |
| | 7 | 8 | 0.158916, 0.5, 0.182169 | 0 | 0.017892 |
| | | | 0.25, 0, 0 | 0 | |
| | | | -0.158915, 0.5, -0.182169 | 0 | |
| | 8 | 16 | 0.158916, 0.5, 0.182169 | 1 | -0.004710 |
| | | | 0.25, 0, 0 | 0 | |
| | | | -0.158915, 0.5, -0.182169 | 0 | |
| | 9 | 8 | 0.158916, 0.5, 0.182169 | 0 | 0.032268 |
| | | | 0.25, 0, 0 | 1 | |
| | | | -0.158915, 0.5, -0.182169 | 0 | |
| | 10 | 16 | 0.158916, 0.5, 0.182169 | 1 | 0.003190 |
| | | | 0.25, 0, 0 | 1 | |
| | | | -0.158915, 0.5, -0.182169 | 0 | |
| | 11 | 8 | 0.158916, 0.5, 0.182169 | 1 | -0.034169 |
| | | | 0.25, 0, 0 | 0 | |
| | | | -0.158915, 0.5, -0.182169 | 1 | |
| | 12 | 8 | 0.158916, 0.5, 0.182169 | 1 | -0.002671 |
| | | | 0.25, 0, 0 | 1 | |
| | | | -0.158915, 0.5, -0.182169 | 1 | |
| | 13 | 8 | 0.341084, 0.817831, 0.5 | 0 | 0.023892 |
| | | | -0.25, 1, 1 | 0 | |
| | | | -0.158915, 0.5, 0.817831 | 0 | |
| | 14 | 16 | 0.341084, 0.817831, 0.5 | 1 | 0.009498 |
| | | | -0.25, 1, 1 | 0 | |
| | | | -0.158915, 0.5, 0.817831 | 0 | |
| | 15 | 8 | 0.341084, 0.817831, 0.5 | 0 | -0.074086 |
| | | | -0.25, 1, 1 | 1 | |
| | | | -0.158915, 0.5, 0.817831 | 0 | |
| | 16 | 16 | 0.341084, 0.817831, 0.5 | 1 | 0.030568 |
| | | | -0.25, 1, 1 | 1 | |
| | | | -0.158915, 0.5, 0.817831 | 0 | |
| | 17 | 8 | 0.341084, 0.817831, 0.5 | 1 | -0.028259 |
| | | | -0.25, 1, 1 | 0 | |
| | | | -0.158915, 0.5, 0.817831 | 1 | |
| | 18 | 8 | 0.341084, 0.817831, 0.5 | 1 | -0.120890 |
| | | | -0.25, 1, 1 | 1 | |
| | | | -0.158915, 0.5, 0.817831 | 1 | |
| | 19 | 8 | 0.341084, 0.817831, 0.5 | 0 | 0.056588 |
| | | | 0.658916, 0.182169, 0.5 | 0 | |
| | | | -0.158915, 0.5, 0.817831 | 0 | |
| | 20 | 16 | 0.341084, 0.817831, 0.5 | 1 | -0.020770 |
| | | | 0.658916, 0.182169, 0.5 | 0 | |
| | | | -0.158915, 0.5, 0.817831 | 0 | |
| | 21 | 8 | 0.341084, 0.817831, 0.5 | 1 | 0.012710 |
| | | | 0.658916, 0.182169, 0.5 | 1 | |
| | | | -0.158915, 0.5, 0.817831 | 0 | |
| | 22 | 8 | 0.341084, 0.817831, 0.5 | 0 | -0.003616 |
| | | | 0.658916, 0.182169, 0.5 | 0 | |
| | | | -0.158915, 0.5, 0.817831 | 1 | |
| | 23 | 16 | 0.341084, 0.817831, 0.5 | 1 | -0.016022 |
| | | | 0.658916, 0.182169, 0.5 | 0 | |
| | | | -0.158915, 0.5, 0.817831 | 1 | |
| | 24 | 8 | 0.341084, 0.817831, 0.5 | 1 | -0.070950 |
| | | | 0.658916, 0.182169, 0.5 | 1 | |
| | | | -0.158915, 0.5, 0.817831 | 1 | |
| | 25 | 16 | 0.158916, 0.5, 0.182169 | 0 | -0.013248 |
| | | | 0.25, 1, 0 | 0 | |
| | | | 0.341084, 0.817831, -0.5 | 0 | |
| | 26 | 16 | 0.158916, 0.5, 0.182169 | 1 | -0.014225 |
| | | | 0.25, 1, 0 | 0 | |
| | | | 0.341084, 0.817831, -0.5 | 0 | |
| | 27 | 16 | 0.158916, 0.5, 0.182169 | 0 | 0.023968 |
| | | | 0.25, 1, 0 | 1 | |
| | | | 0.341084, 0.817831, -0.5 | 0 | |
| | 28 | 16 | 0.158916, 0.5, 0.182169 | 1 | 0.004640 |
| | | | 0.25, 1, 0 | 1 | |
| | | | 0.341084, 0.817831, -0.5 | 0 | |
| | 29 | 16 | 0.158916, 0.5, 0.182169 | 0 | 0.021333 |
| | | | 0.25, 1, 0 | 0 | |
| | | | 0.341084, 0.817831, -0.5 | 1 | |
| | 30 | 16 | 0.158916, 0.5, 0.182169 | 1 | 0.024917 |
| | | | 0.25, 1, 0 | 0 | |
| | | | 0.341084, 0.817831, -0.5 | 1 | |
| | 31 | 16 | 0.158916, 0.5, 0.182169 | 0 | -0.022750 |
| | | | 0.25, 1, 0 | 1 | |
| | | | 0.341084, 0.817831, -0.5 | 1 | |
| | 32 | 16 | 0.158916, 0.5, 0.182169 | 1 | 0.048235 |
| | | | 0.25, 1, 0 | 1 | |
| | | | 0.341084, 0.817831, -0.5 | 1 | |
| | 33 | 16 | 0.158916, 0.5, 0.182169 | 0 | 0.002844 |
| | | | -0.158915, 0.5, -0.182169 | 0 | |
| | | | 0.341084, 0.817831, -0.5 | 0 | |
| | 34 | 16 | 0.158916, 0.5, 0.182169 | 1 | -0.005031 |
| | | | -0.158915, 0.5, -0.182169 | 0 | |
| | | | 0.341084, 0.817831, -0.5 | 0 | |
| | 35 | 16 | 0.158916, 0.5, 0.182169 | 0 | 0.025947 |
| | | | -0.158915, 0.5, -0.182169 | 1 | |
| | | | 0.341084, 0.817831, -0.5 | 0 | |
| | 36 | 16 | 0.158916, 0.5, 0.182169 | 1 | 0.000902 |
| | | | -0.158915, 0.5, -0.182169 | 1 | |
| | | | 0.341084, 0.817831, -0.5 | 0 | |
| | 37 | 16 | 0.158916, 0.5, 0.182169 | 0 | -0.029303 |
| | | | -0.158915, 0.5, -0.182169 | 0 | |
| | | | 0.341084, 0.817831, -0.5 | 1 | |
| | 38 | 16 | 0.158916, 0.5, 0.182169 | 1 | 0.038793 |
| | | | -0.158915, 0.5, -0.182169 | 0 | |
| | | | 0.341084, 0.817831, -0.5 | 1 | |
| | 39 | 16 | 0.158916, 0.5, 0.182169 | 0 | -0.034332 |
| | | | -0.158915, 0.5, -0.182169 | 1 | |
| | | | 0.341084, 0.817831, -0.5 | 1 | |
| | 40 | 16 | 0.158916, 0.5, 0.182169 | 1 | 0.013453 |
| | | | -0.158915, 0.5, -0.182169 | 1 | |
| | | | 0.341084, 0.817831, -0.5 | 1 | |

**Table S5.** Cluster expansion for hexagonal lattice.

| cluster | labels | multiplicity | coordinates | point function index | ECI |
|---|---|---|---|---|---|
| empty | 1 | 1 | | | -0.075874 |
| point | 1 | 6 | 0.5, 1, 0.2363 | 0 | -0.057053 |
| pair | 1 | 12 | 0.5, 0.5, 0.7637 | 0 | 0.072217 |
| | | | 0.5, 0, 0.7637 | 0 | |
| | 2 | 12 | 0.5, 0.5, 0.7637 | 0 | -0.002617 |
| | | | 0, -0.5, 0.7637 | 0 | |
| | 3 | 3 | 0.5, 0.5, 0.2363 | 0 | 0.000181 |
| | | | 0.5, 0.5, -0.2363 | 0 | |
| | 4 | 3 | 0.5, 0.5, 0.7637 | 0 | -0.022962 |
| | | | 0.5, 0.5, 0.2363 | 0 | |
| | 5 | 6 | 0.5, 0.5, 0.7637 | 0 | -0.016601 |
| | | | -0.5, -0.5, 0.7637 | 0 | |
| | 6 | 12 | 1, 0.5, 0.7637 | 0 | 0.022768 |
| | | | 0, -0.5, 0.7637 | 0 | |
| | 7 | 12 | 1, 0.5, 0.2363 | 0 | 0.006583 |
| | | | 0.5, 0, -0.2363 | 0 | |
| | 8 | 12 | 1, 0.5, 0.7637 | 0 | 0.020257 |
| | | | 0.5, 0, 0.2363 | 0 | |
| | 9 | 12 | 0.5, 0.5, 0.2363 | 0 | 0.004224 |
| | | | 0, -0.5, -0.2363 | 0 | |
| | 10 | 12 | 0.5, 0.5, 0.7637 | 0 | -0.00062 |
| | | | 0, -0.5, 0.2363 | 0 | |
| | 11 | 24 | 1, 0.5, 0.7637 | 0 | -0.003249 |
| | | | 0.5, -1, 0.7637 | 0 | |
| | 12 | 6 | 0.5, 0.5, 0.2363 | 0 | -0.01206 |
| | | | -0.5, -0.5, -0.2363 | 0 | |
| | 13 | 12 | 1, 0.5, 0.2363 | 0 | 0.005575 |
| | | | 0, -0.5, -0.2363 | 0 | |
| | 14 | 6 | 0.5, 0.5, 0.7637 | 0 | 0.00895 |
| | | | -0.5, -0.5, 0.2363 | 0 | |
| | 15 | 12 | 1, 0.5, 0.7637 | 0 | -0.006041 |
| | | | 0, -0.5, 0.2363 | 0 | |
| | 16 | 12 | 1, 0.5, 0.7637 | 0 | -0.003634 |
| | | | -0.5, -1, 0.7637 | 0 | |
| | 17 | 24 | 1, 0.5, 0.2363 | 0 | -0.002141 |
| | | | 0.5, -1, -0.2363 | 0 | |
| | 18 | 24 | 1, 0.5, 0.7637 | 0 | -0.003447 |
| | | | 0.5, -1, 0.2363 | 0 | |
| | 19 | 12 | 0.5, 0.5, 0.7637 | 0 | -0.01043 |
| | | | -0.5, -1.5, 0.7637 | 0 | |
| | 20 | 6 | 0.5, 1, 0.7637 | 0 | 0.023681 |
| | | | -0.5, -1, 0.7637 | 0 | |
| | 21 | 12 | 0.5, 0.5, 0.2363 | 0 | 0.008200 |
| | | | 0.5, -1, -0.2363 | 0 | |
| | 22 | 12 | 0.5, 0.5, 0.7637 | 0 | -0.010471 |
| | | | 0.5, -1, 0.2363 | 0 | |
| | 23 | 24 | 0.5, 0.5, 0.7637 | 0 | 0.016202 |
| | | | -1, -1.5, 0.7637 | 0 | |
| | 24 | 6 | 0.5, 0.5, 0.7637 | 0 | 0.00238 |
| | | | 0.5, 0.5, -0.2363 | 0 | |
| | 25 | 12 | 0.5, 0.5, 0.2363 | 0 | 0.00628 |
| | | | -0.5, -1.5, -0.2363 | 0 | |
| | 26 | 6 | 0.5, 1, 0.2363 | 0 | 0.012144 |
| | | | -0.5, -1, -0.2363 | 0 | |
| | 27 | 24 | 0.5, 0.5, 0.7637 | 0 | 0.000153 |
| | | | 0.5, 0, -0.2363 | 0 | |
| triplet | 1 | 4 | 0.5, 0.5, 0.7637 | 0 | 0.022058 |
| | | | 1, 0.5, 0.7637 | 0 | |
| | | | 0.5, 0, 0.7637 | 0 | |
| | 2 | 12 | 0.5, 0.5, 0.7637 | 0 | 0.002521 |
| | | | 0.5, 0, 0.7637 | 0 | |
| | | | 0, -0.5, 0.7637 | 0 | |
| | 3 | 4 | 0.5, 0.5, 0.7637 | 0 | -0.004792 |
| | | | -0.5, 0, 0.7637 | 0 | |
| | | | 0, -0.5, 0.7637 | 0 | |
| | 4 | 24 | 0.5, 0.5, 0.7637 | 0 | 0.003251 |
| | | | 0.5, 0, 0.7637 | 0 | |
| | | | -0.5, -0.5, 0.7637 | 0 | |
| | 5 | 12 | 0.5, 0.5, 0.7637 | 0 | 0.002200 |
| | | | 0.5, -0.5, 0.7637 | 0 | |
| | | | 0.5, -0.5, 0.7637 | 0 | |
| | 6 | 12 | 1, 0.5, 0.7637 | 0 | -0.003050 |
| | | | 0.5, 0, 0.7637 | 0 | |
| | | | 0, -0.5, 0.7637 | 0 | |
| | 7 | 24 | 1, 0.5, 0.7637 | 0 | 0.002428 |
| | | | 0.5, 0.5, 0.7637 | 0 | |
| | | | 0, -0.5, 0.7637 | 0 | |
| | 8 | 24 | 1, 0.5, 0.2363 | 0 | 0.009779 |
| | | | 0.5, 0, 0.2363 | 0 | |
| | | | 0.5, 0, -0.2363 | 0 | |
| | 9 | 12 | 1, 0.5, 0.2363 | 0 | 0.006173 |
| | | | 0.5, 0.5, 0.2363 | 0 | |
| | | | 0.5, 0, -0.2363 | 0 | |
| | 10 | 12 | 1, 0.5, 0.2363 | 0 | -0.001589 |
| | | | 0.5, -0.5, 0.2363 | 0 | |
| | | | 0.5, 0, -0.2363 | 0 | |
| | 11 | 12 | 1, 0.5, 0.2363 | 0 | 0.005335 |
| | | | 0, -0.5, 0.2363 | 0 | |
| | | | 0.5, 0, -0.2363 | 0 | |
| | 12 | 24 | 1, 0.5, 0.7637 | 0 | 0.005542 |
| | | | 0.5, 0, 0.7637 | 0 | |
| | | | 0.5, 0, 0.2363 | 0 | |
| | 13 | 12 | 1, 0.5, 0.7637 | 0 | 0.008181 |
| | | | 0.5, 0.5, 0.7637 | 0 | |
| | | | 0.5, 0, 0.2363 | 0 | |
| | 14 | 12 | 1, 0.5, 0.7637 | 0 | 0.000381 |
| | | | 0.5, -0.5, 0.7637 | 0 | |
| | | | 0.5, 0, 0.2363 | 0 | |
| | 15 | 12 | 1, 0.5, 0.7637 | 0 | 0.004334 |
| | | | 0, -0.5, 0.7637 | 0 | |
| | | | 0.5, 0, 0.2363 | 0 | |
| | 16 | 24 | 0.5, 0.5, 0.2363 | 0 | 0.008194 |
| | | | 0.5, 0, 0.2363 | 0 | |
| | | | 0, -0.5, -0.2363 | 0 | |
| | 17 | 24 | 0.5, 0.5, 0.2363 | 0 | -0.010855 |
| | | | 0, -0.5, 0.2363 | 0 | |
| | | | 0, -0.5, -0.2363 | 0 | |
| | 18 | 12 | 0.5, 0.5, 0.2363 | 0 | -0.011221 |
| | | | -0.5, 0, 0.2363 | 0 | |
| | | | 0, -0.5, -0.2363 | 0 | |
| | 19 | 24 | 0.5, 0.5, 0.2363 | 0 | -0.000823 |
| | | | -0.5, -0.5, 0.2363 | 0 | |
| | | | 0, -0.5, -0.2363 | 0 | |